\documentclass{article}

\usepackage{microtype}
\usepackage{graphicx}
\usepackage{subfigure}
\usepackage{booktabs}
\usepackage{hyperref}

\usepackage{preprint}

\usepackage[utf8]{inputenc}
\usepackage[T1]{fontenc}
\usepackage{multirow}
\usepackage{tcolorbox}
\tcbuselibrary{breakable,skins}
\usepackage{comment}
\usepackage{placeins}
\usepackage{tabularx}
\usepackage{array}
\usepackage{url}
\usepackage{amsfonts}
\usepackage{nicefrac}
\usepackage{xcolor}
\usepackage{xspace}
\usepackage{amsmath}
\usepackage{wrapfig}
\usepackage{listings}

\usepackage{pgfplots}
\usetikzlibrary{svg.path} 
\pgfplotsset{compat=1.17}

\definecolor{evLRU}{HTML}{8A8A85}  
\definecolor{evCX}{HTML}{2A78D6}   
\definecolor{evOE}{HTML}{EB6834}   
\definecolor{evOFF}{HTML}{1BAF7A}  
\definecolor{evON}{HTML}{EDA100}   
\definecolor{evMG}{HTML}{E87BA4}   

\usepackage{enumitem}
\setitemize{noitemsep,topsep=0pt,parsep=0pt,partopsep=0pt}
\setlistdepth{5}
\setenumerate{noitemsep,topsep=0pt,parsep=0pt,partopsep=0pt}
\setdescription{itemsep=0pt,topsep=0pt,parsep=0pt,partopsep=0pt,itemindent=0em,leftmargin=0.5em}

\newenvironment{denseitemize}{
\begin{itemize}[topsep=2pt, partopsep=0pt, leftmargin=1.5em]
  \setlength{\itemsep}{2pt}
  \setlength{\parskip}{0pt}
  \setlength{\parsep}{0pt}
}{\end{itemize}}

\newenvironment{denseenum}{
\begin{enumerate}[topsep=2pt, partopsep=0pt, leftmargin=1.5em]
  \setlength{\itemsep}{2pt}
  \setlength{\parskip}{0pt}
  \setlength{\parsep}{0pt}
}{\end{enumerate}}

\newcommand{\eg}{{\em e.g.}, }
\newcommand{\ie}{{\em i.e.}, }

\newcommand{\vs}{{\em vs.} }

\newcommand{\heading}[1]{\par\addvspace{.5pt}\noindent\textbf{#1}:\enspace\ignorespaces}

\newcommand{\sysname}{Polymorph}
\newcommand{\sys}{\mbox{\sysname}\xspace}
\newcommand{\cacheext}{cache\_ext\xspace}
\newcommand{\fulltitle}{\sysname: Operating System Kernel Optimization through Artificial Intelligence}

\hypersetup{pdftitle={\fulltitle}, pdfauthor={Dean Sysman, Georgios Liargkovas, Tal Zussman, Hasan Zengin, Petra Hedesiu, Hubertus Franke, Tamar Eilam, Kostis Kaffes, Asaf Cidon}}
\papertitlerunning{\fulltitle}

\begin{document}
\twocolumn[
\papertitle{\fulltitle}

\begin{paperauthorlist}
\paperauthor{Dean Sysman}{cu}
\paperauthor{Georgios Liargkovas}{cu}
\paperauthor{Tal Zussman}{cu}
\paperauthor{Hasan Zengin}{cu}
\paperauthor{Petra Hedesiu}{cu}
\paperauthor{Hubertus Franke}{ibm}
\paperauthor{Tamar Eilam}{ibm}
\paperauthor{Kostis Kaffes}{cu}
\paperauthor{Asaf Cidon}{cu}
\end{paperauthorlist}
\paperaffiliation{cu}{Columbia University}
\paperaffiliation{ibm}{IBM}
\papercorrespondingauthor{Dean Sysman}{ds4597@columbia.edu}
\paperkeywords{Machine Learning, Operating Systems, Page Cache, Online Adaptation}
\vskip 0.3in
\begin{abstract}
While popular operating systems such as Linux and Windows were originally designed as monolithic kernels, an increasing number of kernel subsystems have become extensible using eBPF. Users can now define custom policies for essential kernel mechanisms, such as the process scheduler, page cache, and network stack, without recompiling the kernel. These mechanisms give developers significant freedom to fit the kernel to their applications' workloads, but the question of which policy to deploy remains open. We study how to use agentic AI to automate kernel policy development, selection, and deployment, focusing on the page cache and the process scheduler as two representative subsystems. We show that existing approaches fall short: evolutionary and agentic search runs offline, is relatively expensive, and often fails to generalize to online workloads, while online knob tuners are encumbered by the knobs' lack of expressivity. In contrast, we propose \sys, which learns online: an AI agent writes a generalized, granular policy once, and a user mode tuner adjusts its parameters during runtime, with no LLM in the loop. For the page cache, \sys improves throughput over Linux's LRU by 8.4\% on unseen data, nearly twice the gain of the best existing policy, while adapting to workload changes at runtime and costing only 16M tokens once, versus 125--240M tokens per workload for agentic offline search. For the process scheduler, it improves on Linux's default by 2.75$\times$ across six workloads, versus 2.27$\times$ for the best fixed \texttt{sched\_ext} policy.
\end{abstract}
]
\printAffiliationsAndNotice{}
\section{Introduction}
\label{sec:introduction}

Legacy operating systems, such as Linux and Windows, were designed as monolithic OSes with fixed, predefined interfaces and data structures, which give applications limited ways to control their behavior. For example, the Linux page cache implements an approximation of the Least-Recently-Used (LRU) policy and lets applications tweak it with ``hints'' through the \texttt{madvise()} and \texttt{fadvise()} interfaces. However, prior work has shown that these hints may have little impact on the behavior of the underlying policy~\cite{cacheext,p2cache}.

In recent years, mainline operating systems have become more \emph{extensible} through the eBPF framework, which allows applications to safely run custom, verified code in the kernel~\cite{linux_bpf_docs}. Beyond its original uses in networking, security, and tracing, eBPF now extends subsystems such as scheduling, memory management, and storage. For example, \texttt{sched\_ext} and \cacheext allow applications to load custom policies for process scheduling and page cache eviction, respectively~\cite{sched_ext,cacheext}.

Custom policies matter because no single policy suits every application, and Linux alone makes hundreds of such policy decisions, most of which cannot be changed without rebuilding the kernel~\cite{qin2026policies}. LLM-based code generation now makes new policies cheap to write~\cite{novikov2025alphaevolve,cheng2025barbarians}, which shifts the bottleneck to designing and deploying the right one. This is difficult: the design space is broad, it is hard to predict how a workload will behave under a given policy, and workloads change while they run, so a policy that suits one phase can hurt the next.

In this paper, we explore how AI and agentic workflows can help develop and select custom OS policies, and we investigate several approaches for using AI and agents to design, implement, and deploy them. We find that the most effective approach is to use AI once to build a policy that adapts itself online as the workload changes, which performs consistently on unseen data and costs less than generating policies offline (\S\ref{sec:eval-online}--\ref{sec:eval-sched}).

\paragraph{Offline agentic iterations.} Recent evolutionary frameworks, such as OpenEvolve~\cite{openevolve} and SkyDiscover~\cite{skydiscover2026}, iteratively generate and evaluate algorithms that optimize a particular objective (e.g., throughput or cache hit rate) on an offline trace. We use these tools to generate eBPF code, verify it, and load it into the kernel as a novel page cache or scheduling policy (\S\ref{subsec:evolutionary_search}). We find that these policies generally yield poor results: they are expensive and do not generalize to online, unseen data. Such frameworks optimize only the objective function, while treating the system (an operating system running applications with particular workloads) as a black box. We therefore incorporate ``white box'' information about the workload, such as OS-level I/O traces, into the feedback loop (\S\ref{subsec:offline_learning}). This results in significant improvements over ``vanilla'' OpenEvolve, but the core problems of the offline approach remain: it is expensive, requiring many iterations on offline data, and it fails to generalize to new, unseen workloads.

\paragraph{Online parameter search.} OS configuration tuners search the parameters that the kernel exposes~\cite{alipourfard2017cherrypick,van2017automatic,wang2018understanding}, increasingly with LLMs~\cite{GPTuner2024,chen2024autoos}. TuxBot~\cite{liargkovas2026tuxbot} tunes them online, so it is lightweight and can react to workload changes, but it can only choose among behaviors that existing policies implement. When TuxBot tunes existing Linux knobs such as \texttt{fadvise()} (\S\ref{sec:knob-tuning}), it results in coarse effects, such as disabling readahead entirely.

\paragraph{\sys: AI-driven online policies.} We propose \sys, a novel approach to AI-driven OS customization that combines the benefits of both approaches (\S\ref{sec:design}). \sys uses AI agents to create new eBPF-based policies that generalize and react to workload changes, without invoking LLMs at runtime.

We apply \sys to two OS policies, the page cache and the process scheduler. For the page cache, on 11 production traces and data that no approach saw during generation, \sys improves throughput over Linux's LRU by 8.4\% on average, versus 4.2\% for the best existing \cacheext policy. Offline-generated policies gained at most 3\%. Generating a policy with \sys costs about 16M tokens once, versus 125--240M tokens per workload for agentic offline search. For the process scheduler, \sys improves on Linux's default scheduler by 2.75$\times$ on average across six workloads, versus 2.27$\times$ for the best fixed \texttt{sched\_ext} policy, and adapts to workload changes within 2.5 seconds.

We make the following contributions:
\begin{denseitemize}
    \item We show that offline AI-generated OS policies are expensive and do not generalize to unseen workloads, while tuning existing knobs online is too limited (\S\ref{subsec:evolutionary_search}--\ref{sec:knob-tuning}).
    \item We design \sys, which uses an AI agent once to write an eBPF policy with tunable parameters, and a non-LLM userspace tuner that adapts them online (\S\ref{sec:design}).
    \item We evaluate \sys on 17 workloads across the page cache (11) and the process scheduler (6) (\S\ref{sec:evaluation}).
\end{denseitemize}

\section{Background}
\label{sec:background}

\subsection{Customizing OS Policies with eBPF}

eBPF (Extended Berkeley Packet Filter) is a Linux framework for safely running custom code in the kernel with the help of a verifier to ensure safety. It was initially envisioned as a way to filter and forward packets in kernel space without having to copy them into userspace. Over the past decade, it has expanded from this initial use case to many domains, including tracing, memory and storage~\cite{p2cache,xrp,fetchbpf,bpf_fault}, and scheduling~\cite{syrup,ghost}.
Recently, frameworks like \texttt{sched\_ext} and \cacheext allow applications to load custom eBPF policies for scheduling and page cache eviction, respectively, without having to recompile the kernel~\cite{sched_ext,cacheext}. %
Such policies can be switched out or adjusted live by using shared memory regions between userspace and eBPF programs. These frameworks turn hard-coded kernel mechanisms into flexible, customizable subsystems.
\vspace{-6pt}

\paragraph{Page cache.}
By default, Linux approximates LRU with an active and an inactive list per cgroup. Newly added pages enter the inactive list and are promoted to the active list when accessed again. Reclaim evicts pages from the tail of the inactive list and demotes pages from the active list to keep the two balanced. MGLRU, merged in Linux 6.1, generalizes this to multiple generations ordered by access recency. \cacheext exposes eviction to eBPF: a policy implements hooks that run when pages are added to, accessed in, or removed from the page cache, and when the kernel needs eviction candidates. The policy tracks pages in its own eviction lists and returns eviction candidates by iterating over these lists with a callback. Policies are attached per cgroup, so different applications can run different policies. \cacheext ships with several such policies, including LFU, LHD~\cite{211237} and S3-FIFO~\cite{yang2023s3fifo}, which we use as baselines in our experiments.

\paragraph{Scheduling.}
Linux's default scheduler for normal tasks is EEVDF (Earliest Eligible Virtual Deadline First), which replaced CFS in Linux 6.6. EEVDF tracks each task's virtual runtime and runs the eligible task with the earliest virtual deadline, which is derived from the task's weight and time slice. \texttt{sched\_ext} adds a new scheduling class whose operations, such as selecting a CPU and enqueueing and dispatching tasks, are implemented in eBPF. If the eBPF scheduler errors or stalls runnable tasks, the kernel unloads it and falls back to the default scheduler.

\subsection{OS Policy Tuning}
Besides replacing a policy, users can adjust existing OS policies.
Linux exposes over 140 performance-related runtime parameters (or knobs) through sysctl, sysfs, and debugfs.
These include scheduler time slices, CPU frequency limits, and readahead sizes, along with hints such as \texttt{fadvise()}.
Xkernel further expands this knob set by making OS constants tunable on a running kernel~\cite{chen2025xkernel}.

Traditionally, performance engineers chose knob values through manual experimentation and rules of thumb. Automated tuners replace that process with a search over knob values against a measured objective, typically driven by Bayesian optimization~\cite{curino2020mlos} or reinforcement learning~\cite{karthikeyan2023selftune}. 
Recently, TuxBot replaced the optimizer with an LLM that reasons about knob semantics and system telemetry to tune these parameters on a live host~\cite{liargkovas2026tuxbot, liargkovas2025an}.

\section{Motivation: Why Do We Need Policy Customization?}
\label{sec:motivation}

OS policy choice can have a dramatic effect on application performance~\cite{cacheext-journal,p2cache,syrup}. To demonstrate this, we vary the policy of two OS mechanisms, the page cache and the CPU scheduler, and measure how real workloads respond.

\vspace{-0.5em}
\paragraph{Page cache.} For the page cache, we run a collection of 11 storage traces from Twitter~\cite{258935} on a key-value store (LevelDB), and Meta~\cite{MetaPlatforms2026} directly through the filesystem, while varying the OS page cache eviction policy. We try five different policies. The first two are hard-coded Linux page cache policies: the LRU (least-recently used) approximate policy, and the more recently introduced MGLRU (multi-generational least-recently used) policy, which ships on some Linux distributions. We also test three eBPF-based policies from the \cacheext paper: LFU, LHD~\cite{211237}) and S3FIFO~\cite{yang2023s3fifo}.

\begin{figure}[t]
  \centering
  \resizebox{\columnwidth}{!}{\input{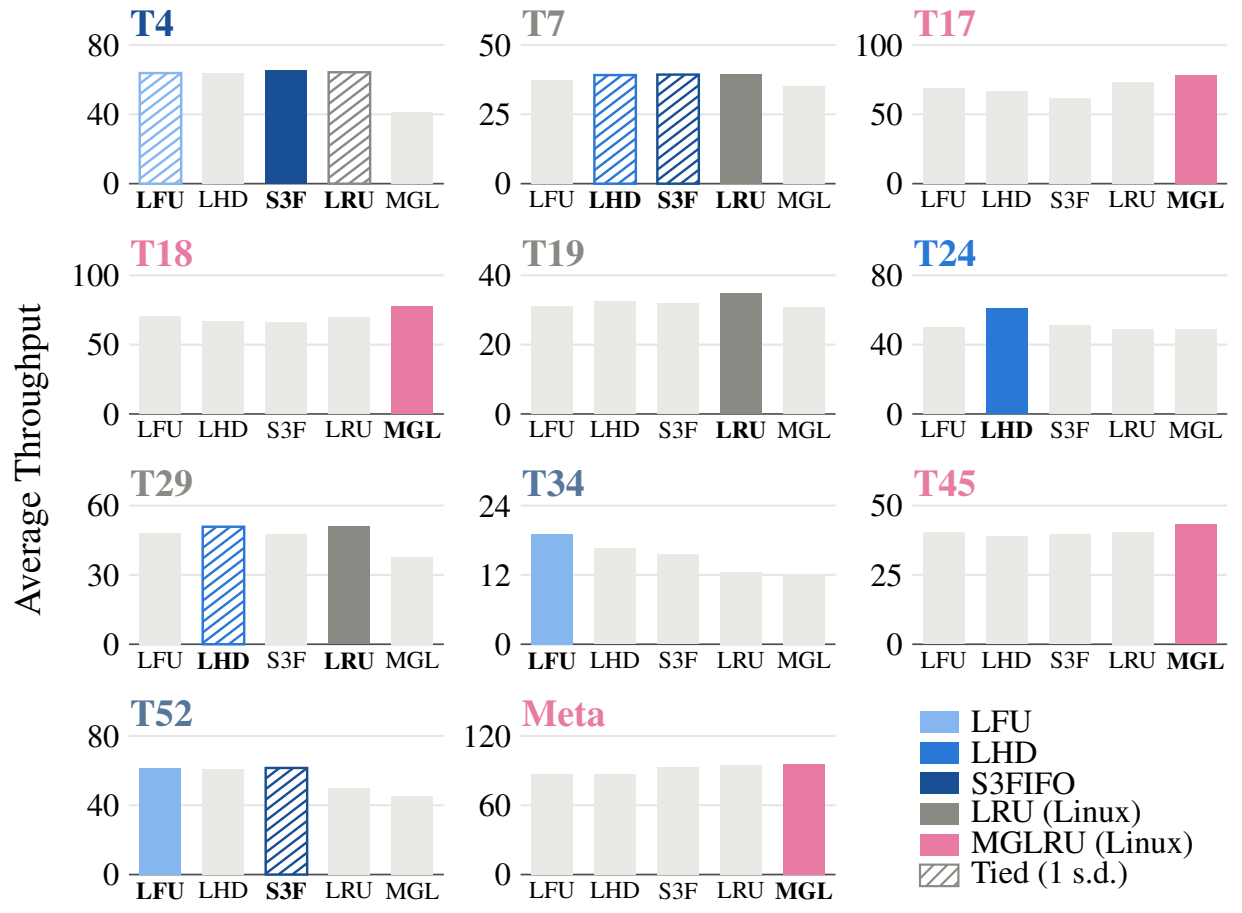}}
  \vspace{-1em}
  \caption{LevelDB (Twitter) and filesystem (Meta) throughput with different page cache policies.}
  \label{fig:long_wins_ties}
  \vspace{-1em}
\end{figure}

 We run each trace three times on each cache policy. Our experimental setup is described in \S\ref{sec:evaluation}.
The choice of page cache policy alone changes end-to-end throughput substantially (Figure~\ref{fig:long_wins_ties}). For example, on Twitter cluster 34 (``T34'' in the figure), LevelDB's throughput is about 50\% higher with \cacheext's LFU, the best policy, than with the kernel's MGLRU, the worst. Moreover, no single policy is best on all workloads, or even on a majority of them, and the winner follows no clear pattern.

\vspace{-0.5em}
\paragraph{Process scheduler.} Unsurprisingly, the OS process scheduler policy can also significantly affect the performance of individual applications. We ran four workloads with Linux's default scheduler, EEVDF, and with four \texttt{sched\_ext}~\cite{sched_ext} policies, and measured throughput for batch work and 99th-percentile (p99) latency for several services (Table~\ref{tab:sched_motivation}). Each workload runs on 11 hardware threads. The CPU-bound workload runs 11 always-busy batch threads alongside 44 threads that compute briefly and sleep, and the other three are services driven at a fixed request rate: a PostgreSQL database under Sysbench OLTP~\cite{kopytov2024sysbench}, alone or with 11 batch threads, and the Xapian search engine~\cite{kasture2016tailbench}. The policies differ in a few ways. \texttt{sched\_ext's} default policy runs all tasks FIFO from one global queue with 20\,ms time slices. \emph{Long-slice} favors throughput: it gives every task a 20\,ms slice, and a task that wakes up never moves ahead of tasks that are already waiting. \emph{Wake-boost} favors responsiveness: it uses 5\,ms slices and lets a newly woken task pass up to 16\,ms of waiting work. \emph{Per-socket} keeps a separate queue for each socket and uses 5\,ms slices.

\begin{table}[t]
  \centering
  \resizebox{\columnwidth}{!}{%
  \begin{tabular}{@{}lrrrr@{}}
    \toprule
    Policy & CPU-bound (k/s\,$\uparrow$) & OLTP + batch (ms\,$\downarrow$) & OLTP (ms\,$\downarrow$) & Xapian (ms\,$\downarrow$) \\
    \midrule
    EEVDF (default) & 7.6 & 17.9 & 8.90 & 57.2 \\
    \texttt{sched\_ext} default & 10.1 & 10,159 & 3.25 & \textbf{14.7} \\
    Long-slice & \textbf{14.3} & 4,437 & 2.81 & 15.8 \\
    Wake-boost & 4.5 & \textbf{4.3} & \textbf{2.76} & 15.4 \\
    Per-socket & 10.5 & 36.2 & 2.86 & 15.0 \\
    \bottomrule
  \end{tabular}}
  \vspace{-1em}
  \caption{Batch throughput (CPU-bound) and p99 latency (others) of four workloads under EEVDF and four \texttt{sched\_ext} policies. Median of three runs. The best result is in \textbf{bold}.}
  \label{tab:sched_motivation}
  \vspace{-1em}
\end{table}

The best policy for one workload can be the worst for another. On a CPU-bound workload, long-slice nearly doubles batch throughput over EEVDF, because batch threads keep their CPUs for long stretches. Wake-boost performs worst on this workload, 41\% below EEVDF, because threads that wake up often jump ahead of the batch threads. The ranking flips when a PostgreSQL database running Sysbench OLTP~\cite{kopytov2024sysbench} shares the CPUs with batch threads. With long-slice, database processes wait behind 20\,ms batch slices, the database falls behind, and P99 latency grows from 18\,ms with EEVDF to 4.4\,s. Wake-boost lets them jump ahead and cuts P99 latency to 4.3\,ms, 4$\times$ below EEVDF.

For services that run alone, such as the same database or the Xapian search engine, every \texttt{sched\_ext} policy lowers P99 latency by 3--4$\times$, likely because it serves tasks in arrival order rather than sharing each CPU among them. On Xapian, \texttt{sched\_ext's} default policy and per-socket do best.

These observations point to a recurring pattern: OS policies significantly impact workload performance, and a policy that suits one workload can severely hurt another. No single policy can therefore be deployed once for all applications. Instead, operators need a way to choose and adapt OS policies for each application and workload.

\section{Existing AI-based Approaches}

\begin{figure}[t]
  \centering
  \resizebox{0.8\linewidth}{!}{\begingroup%
\definecolor{archfill}{gray}{0.92}%
\definecolor{archmap}{gray}{0.80}%
\definecolor{archline}{gray}{0.15}%
\definecolor{archmuted}{gray}{0.35}%
\def\fs#1{\fontfamily{phv}\fontsize{\dimexpr #1pt*92/100\relax}%
  {\dimexpr #1pt*110/100\relax}\selectfont}%
\def\note#1{{\fs{7}\color{archmuted}#1}}%
\input{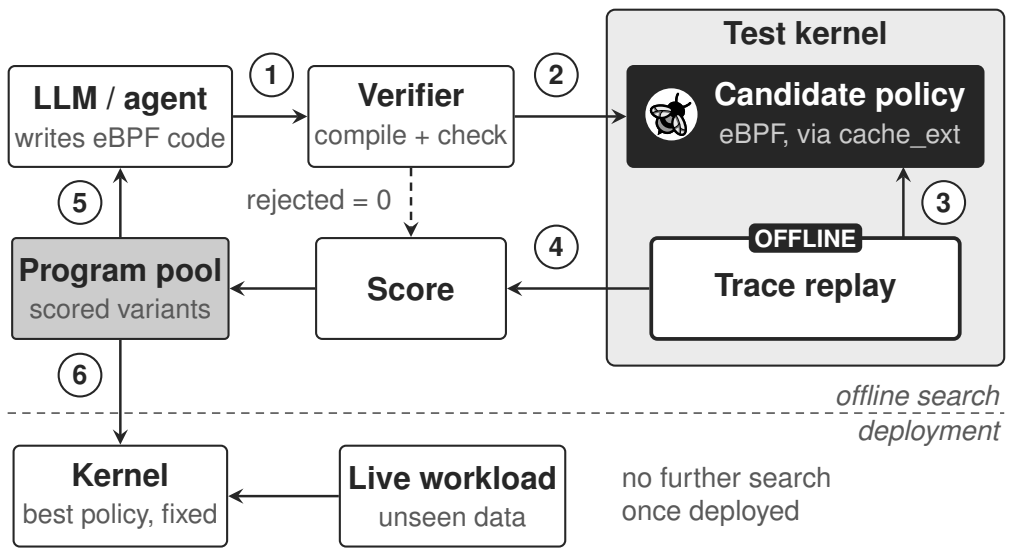}%
\fontfamily{phv}\selectfont%
\resizebox{\linewidth}{!}{%
\begin{tikzpicture}[
  archline, line width=0.5pt, >=stealth,
  container/.style={draw=archline, fill=archfill, rounded corners=2pt},
  comp/.style={draw=archline, fill=white, rounded corners=1.5pt, align=center,
               inner sep=1.5pt, font=\fs{8.2}, minimum height=0.85cm},
  map/.style={draw=archline, fill=archmap, rounded corners=1pt, align=center,
              inner sep=1.5pt, minimum height=0.85cm, font=\fs{8.2}},
  sub/.style={font=\fs{7.2}, text=archmuted},
  stepno/.style={circle, draw=archline, fill=white, inner sep=0.4pt,
               minimum size=0.36cm, font=\fs{7}\bfseries},
  flow/.style={->, line width=0.6pt},
  pill/.style={rounded corners=1.5pt, inner sep=1.4pt, fill=archline, text=white,
               font=\fs{6.2}\bfseries},
  input/.style={line width=0.9pt},
  iface/.style={fill=archline, text=white},
]

\draw[archmuted, dash pattern=on 2pt off 1.5pt, line width=0.4pt] (0,1.55) -- (8.4,1.55);
\node[anchor=south east, sub, font=\fs{7.2}\itshape, inner sep=1pt] at (8.4,1.57) {offline search};
\node[anchor=north east, sub, font=\fs{7.2}\itshape, inner sep=1pt] at (8.4,1.53) {deployment};

\node[comp, minimum width=1.8cm] (llm) at (0.95,4.05)
  {\textbf{LLM / agent}\\\note{writes eBPF code}};
\node[comp, minimum width=1.6cm] (ver) at (3.4,4.05)
  {\textbf{Verifier}\\\note{compile + check}};
\node[map, minimum width=1.8cm] (pool) at (0.95,2.6)
  {\textbf{Program pool}\\\note{scored variants}};
\node[comp, minimum width=1.6cm] (score) at (3.4,2.6)
  {\textbf{Score}};

\draw[container] (5.05,1.95) rectangle (8.4,4.95);
\node[anchor=north, font=\fs{8.2}\bfseries, inner sep=0pt] at (6.725,4.86) {Test kernel};
\node[comp, iface, minimum width=3.0cm] (pol) at (6.725,4.05)
  {\raisebox{\dimexpr -0.5\height+2.5pt\relax}{\ebpfbee}\hspace{4pt}\begin{tabular}{@{}c@{}}\textbf{Candidate policy}\\{\fs{7}\color{archmap}eBPF, via \cacheext}\end{tabular}};
\node[comp, input, minimum width=2.6cm] (trace) at (6.725,2.6)
  {\textbf{Trace replay}};
\node[pill] at (6.725,3.025) {OFFLINE};

\node[comp, minimum width=1.8cm] (prod) at (0.95,0.85)
  {\textbf{Kernel}\\\note{best policy, fixed}};
\node[comp, minimum width=1.9cm] (live) at (3.75,0.85)
  {\textbf{Live workload}\\\note{unseen data}};
\node[sub, anchor=west, align=left] at (5.05,0.85) {no further search\\once deployed};

\draw[flow] (llm) -- (ver);
\draw[flow] (ver) -- (pol);
\draw[flow] (7.55,3.025) -- (7.55,3.625);
\draw[flow] (trace) -- (score);
\draw[flow] (score) -- (pool);
\draw[flow] (pool) -- (llm);
\draw[flow, dash pattern=on 2pt off 1.5pt] (ver) -- node[sub, left=1pt] {rejected = 0} (score);
\draw[flow] (pool) -- (prod);
\draw[flow] (live) -- (prod);

\node[stepno] at (2.225,4.4) {1};
\node[stepno] at (4.625,4.4) {2};
\node[stepno] at (7.89,3.325) {3};
\node[stepno] at (4.625,2.95) {4};
\node[stepno] at (0.61,3.325) {5};
\node[stepno] at (0.61,1.87) {6};

\end{tikzpicture}}%
\endgroup%
}
  \vspace{-6pt}
  \caption{Offline policy generation with OpenEvolve.}
  \label{fig:offline_search_loop}
  \vspace{-1em}
\end{figure}

In this section, we explore two offline approaches (evolutionary methods and iteration with agentic coding) and one online approach (online parameter tuning) for OS policy development, and demonstrate why they cannot efficiently adapt OS policies to online workloads.

\subsection{Evolutionary Search Using LLMs}
\label{subsec:evolutionary_search}

Recent work~\cite{cheng2025barbarians,AGIsystems,policysmith} applies evolutionary search-style algorithms inspired by AlphaEvolve~\cite{novikov2025alphaevolve} to systems contexts. Such systems take an offline trace or use a synthetic simulator, and generate a policy to try to optimize the objective (\eg throughput, latency, cache hit rate), and based on the achieved score create a new iteration of the policy. These frameworks generally have LLMs generate different hypotheses, which are tested on real data, validated, and then discarded or merged.

In order to test this approach, we use OpenEvolve~\cite{openevolve}, an open source variant of AlphaEvolve, and generate custom eBPF page cache policies for \cacheext. We also tested the SkyDiscover~\cite{skydiscover2026} framework, but it failed to generate eBPF code that passed the verifier.

\begin{figure}
    \centering
    \input{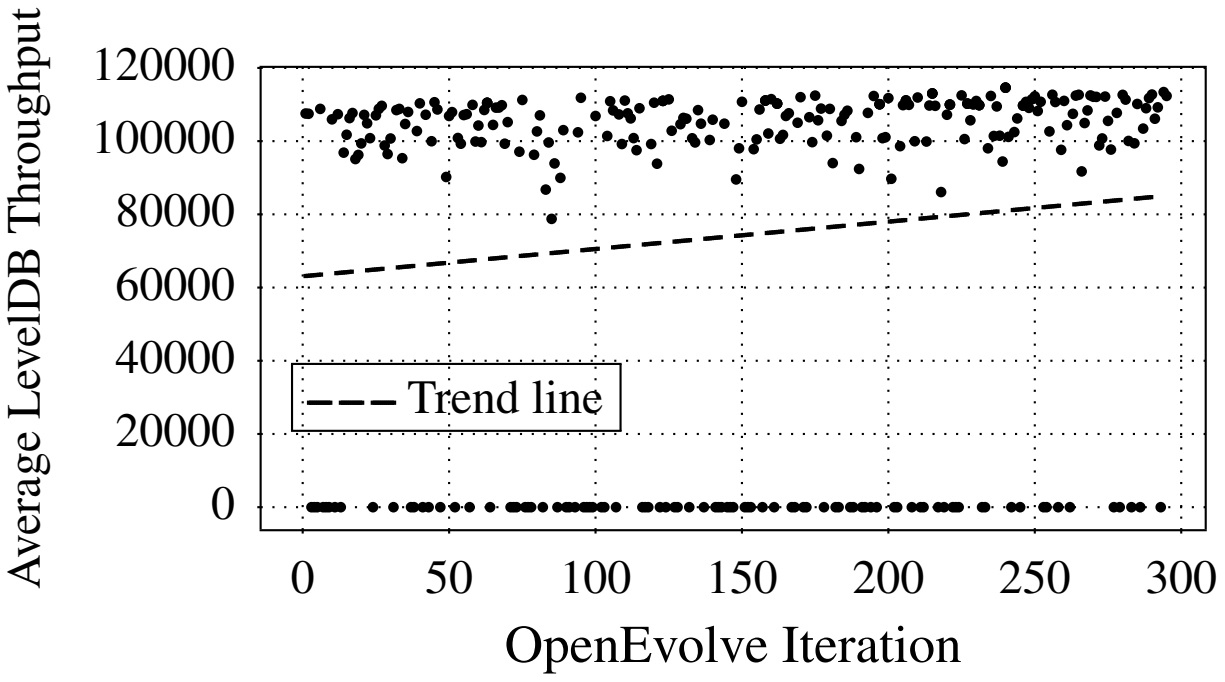}
    \vspace{-1em}
    \caption{Throughput using OpenEvolve on Twitter cluster 17.}
    \label{fig:evolutionary-search-iterations}
    \vspace{-1em}
\end{figure}

Figure~\ref{fig:offline_search_loop} shows how we deployed OpenEvolve in the context of eBPF-based policies. We use an LLM (Gemini 3.1 Pro, which we found performed best with OpenEvolve) to generate an eBPF caching policy~\textcircled{\scriptsize 1}. The policy is then compiled and must pass the eBPF verifier~\textcircled{\scriptsize 2}. If it fails, the LLM tries to create a new eBPF policy. If it succeeds, it is loaded into the kernel, and we then measure how well it performs on a short online trace, which is a random segment from a cache trace~\textcircled{\scriptsize 3}. We first ``warm up'' each run for 45 seconds, and then measure the ``score'' (\ie LevelDB's average throughput) over the next 4 minutes of the trace~\textcircled{\scriptsize 4}.
We save the eBPF program into a ``program pool'' alongside its score, and OpenEvolve continues to iterate. At the end of the run, we load the highest scoring eBPF policy into the kernel, and use it for the rest of the trace on unseen data~\textcircled{\scriptsize 5}.

Figure~\ref{fig:evolutionary-search-iterations} shows how the score improves over successive iterations. Interestingly, a large portion of the attempts fail to compile or result in an error (indicated by throughput of 0). In addition, the improvement is not monotonic -- OpenEvolve tests different hypotheses, some of which fail. Nevertheless, the overall trend is positive: OpenEvolve achieves higher throughput over time on the offline trace.

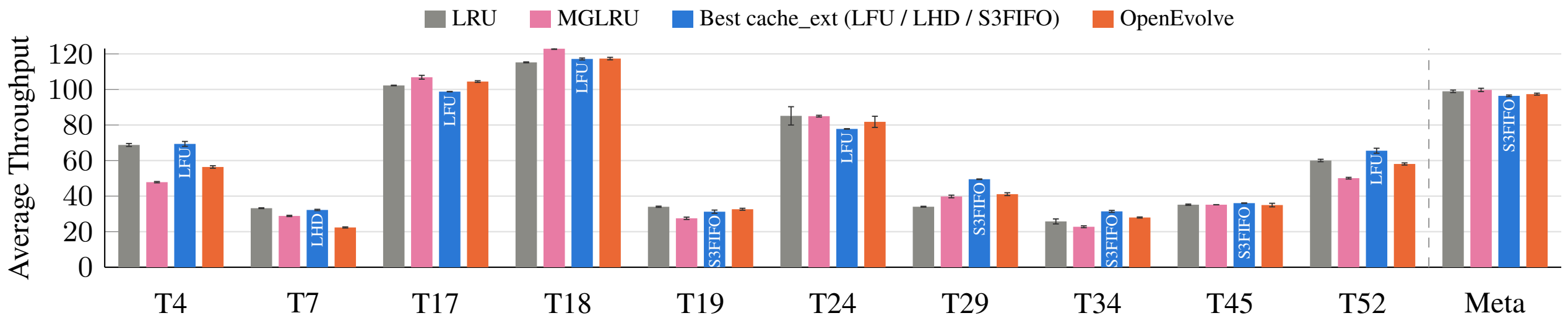
\begin{figure*}[t]
  \centering
  \begin{tikzpicture}
\begin{axis}[
    width=\textwidth, height=3.92cm,
    ybar, bar width=6.5pt,
    ymin=0, ymax=123, ytick distance=20,
    xmin=-0.5, xmax=10.5,
    xtick={0,...,10},
    xticklabels={T4,T7,T17,T18,T19,T24,T29,T34,T45,T52,Meta},
    xtick style={draw=none},
    tick label style={font=\footnotesize},
    label style={font=\footnotesize},
    ylabel={Average Throughput},
    ylabel shift=-3pt,
    ymajorgrids, grid style={black!12},
    axis x line*=bottom, axis y line*=left,
    every axis plot/.append style={draw=none},
    error bars/y dir=both, error bars/y explicit,
    error bars/error bar style={black!80, line width=0.35pt},
    error bars/error mark options={rotate=90, mark size=0.9pt, black!80, line width=0.35pt},
    every node near coord/.append style={
        rotate=90, anchor=east, font=\fontsize{5}{6}\selectfont, text=white, inner sep=0.8pt},
    legend style={draw=none, fill=none, font=\scriptsize, at={(0.5,1.02)}, anchor=south,
                  legend columns=4, legend cell align=left,
                  /tikz/every even column/.append style={column sep=8pt}},
    legend image code/.code={\fill[#1] (0cm,-0.08cm) rectangle (0.22cm,0.1cm);},
]
\addplot+[fill=evLRU] coordinates {
    (0,68.78) +- (0,0.82)
    (1,33.21) +- (0,0.22)
    (2,102.24) +- (0,0.16)
    (3,115.23) +- (0,0.25)
    (4,34.01) +- (0,0.34)
    (5,85.14) +- (0,5.15)
    (6,34.04) +- (0,0.24)
    (7,25.77) +- (0,1.41)
    (8,35.18) +- (0,0.35)
    (9,60.03) +- (0,0.69)
    (10,98.96) +- (0,0.70)
};
\addlegendentry{LRU}
\addplot+[fill=evMG] coordinates {
    (0,47.85) +- (0,0.36)
    (1,28.85) +- (0,0.35)
    (2,106.86) +- (0,1.08)
    (3,122.90) +- (0,0.35)
    (4,27.53) +- (0,0.65)
    (5,84.97) +- (0,0.48)
    (6,39.81) +- (0,0.73)
    (7,22.76) +- (0,0.52)
    (8,35.18) +- (0,0.03)
    (9,50.09) +- (0,0.47)
    (10,99.74) +- (0,0.93)
};
\addlegendentry{MGLRU}
\addplot+[fill=evCX, nodes near coords, point meta=explicit symbolic] coordinates {
    (0,69.37) +- (0,1.38) [LFU]
    (1,32.22) +- (0,0.36) [LHD]
    (2,98.77) +- (0,0.10) [LFU]
    (3,117.15) +- (0,0.54) [LFU]
    (4,31.24) +- (0,0.91) [S3FIFO]
    (5,77.80) +- (0,0.15) [LFU]
    (6,49.50) +- (0,0.16) [S3FIFO]
    (7,31.42) +- (0,0.56) [S3FIFO]
    (8,36.06) +- (0,0.12) [S3FIFO]
    (9,65.54) +- (0,1.40) [LFU]
    (10,96.40) +- (0,0.47) [S3FIFO]
};
\addlegendentry{Best \cacheext (LFU / LHD / S3FIFO)}
\addplot+[fill=evOE] coordinates {
    (0,56.39) +- (0,0.69)
    (1,22.38) +- (0,0.31)
    (2,104.44) +- (0,0.47)
    (3,117.37) +- (0,0.69)
    (4,32.62) +- (0,0.55)
    (5,81.79) +- (0,3.14)
    (6,41.10) +- (0,0.80)
    (7,27.98) +- (0,0.29)
    (8,34.97) +- (0,1.00)
    (9,58.09) +- (0,0.60)
    (10,97.36) +- (0,0.54)
};
\addlegendentry{OpenEvolve}
\draw[black!40, dashed] (axis cs:9.5,0) -- (axis cs:9.5,123);
\end{axis}
\end{tikzpicture}
  \vspace{-1em}
  \caption{Average throughput for short runs on offline traces on the cache traces, comparing OpenEvolve with the \cacheext and kernel policies.}

  \label{fig:short_oe_throughput_by_cluster}
  \vspace{-1em}
\end{figure*}

However, in our experiments, OpenEvolve did not produce significant throughput increases over the default kernel and \cacheext policies, even on its short offline traces. Figure~\ref{fig:short_oe_throughput_by_cluster} plots the average throughput comparing the best performing \cacheext policy among LFU, LHD and S3FIFO, to the kernel's LRU, MGLRU and OpenEvolve. OpenEvolve did not achieve the highest performing throughput in any cluster, and it only managed to achieve higher throughput than the default \cacheext policies in Twitter cluster 24. When examining the policies produced by OpenEvolve, they were extremely overfitted, with various hard coded parameters that seem to be designed to perform well on the particular last trace it was run on.
This approach failed to yield \emph{generalizable} policies. %

\subsection{White Box Agentic Iterations}
\label{subsec:offline_learning}

\begin{figure*}[t]
  \centering
  \begin{tikzpicture}
  \definecolor{agLRU}{HTML}{8A8A85}%
  \definecolor{agMG}{HTML}{E87BA4}%
  \definecolor{agCX}{HTML}{2A78D6}%
  \definecolor{agOE}{HTML}{EB6834}%
  \definecolor{agReq}{HTML}{1BAF7A}%
  \definecolor{agIO}{HTML}{6250D6}%
\begin{axis}[
    width=\textwidth, height=3.92cm,
    ybar, bar width=3.9pt,
    ybar=0.5pt,
    ymin=0, ymax=123, ytick distance=20,
    xmin=-0.5, xmax=10.5,
    xtick={0,...,10},
    xticklabels={T4,T7,T17,T18,T19,T24,T29,T34,T45,T52,Meta},
    xtick style={draw=none},
    tick label style={font=\footnotesize},
    label style={font=\footnotesize},
    ylabel={Average Throughput},
    ylabel shift=-3pt,
    ymajorgrids, grid style={black!12},
    axis x line*=bottom, axis y line*=left,
    every axis plot/.append style={draw=none},
    error bars/y dir=both, error bars/y explicit,
    error bars/error bar style={black!80, line width=0.35pt},
    error bars/error mark options={rotate=90, mark size=0.9pt, black!80, line width=0.35pt},
    every node near coord/.append style={
        rotate=90, anchor=east, font=\fontsize{5}{6}\selectfont, text=white, inner sep=0.8pt},
    legend style={draw=none, fill=none, font=\scriptsize, at={(0.5,1.02)}, anchor=south,
                  legend columns=6, legend cell align=left,
                  /tikz/every even column/.append style={column sep=4pt}},
    legend image code/.code={\fill[#1] (0cm,-0.08cm) rectangle (0.22cm,0.1cm);},
]
\addplot+[fill=agLRU] coordinates {
    (0,68.78) +- (0,0.82)
    (1,33.21) +- (0,0.22)
    (2,102.24) +- (0,0.16)
    (3,115.23) +- (0,0.25)
    (4,34.01) +- (0,0.34)
    (5,85.14) +- (0,5.15)
    (6,34.04) +- (0,0.24)
    (7,25.77) +- (0,1.41)
    (8,35.18) +- (0,0.35)
    (9,60.03) +- (0,0.69)
};
\addlegendentry{LRU}
\addplot+[fill=agMG] coordinates {
    (0,47.85) +- (0,0.36)
    (1,28.85) +- (0,0.35)
    (2,106.86) +- (0,1.08)
    (3,122.90) +- (0,0.35)
    (4,27.53) +- (0,0.65)
    (5,84.97) +- (0,0.48)
    (6,39.81) +- (0,0.73)
    (7,22.76) +- (0,0.52)
    (8,35.18) +- (0,0.03)
    (9,50.09) +- (0,0.47)
};
\addlegendentry{MGLRU}
\addplot+[fill=agCX, nodes near coords, point meta=explicit symbolic] coordinates {
    (0,69.37) +- (0,1.38) [LFU]
    (1,32.22) +- (0,0.36) [LHD]
    (2,98.77) +- (0,0.10) [LFU]
    (3,117.15) +- (0,0.54) [LFU]
    (4,31.24) +- (0,0.91) [S3FIFO]
    (5,77.80) +- (0,0.15) [LFU]
    (6,49.50) +- (0,0.16) [S3FIFO]
    (7,31.42) +- (0,0.56) [S3FIFO]
    (8,36.06) +- (0,0.12) [S3FIFO]
    (9,65.54) +- (0,1.40) [LFU]
};
\addlegendentry{Best \cacheext (LFU / LHD / S3FIFO)}
\addplot+[fill=agOE] coordinates {
    (0,56.39) +- (0,0.69)
    (1,22.38) +- (0,0.31)
    (2,104.44) +- (0,0.47)
    (3,117.37) +- (0,0.69)
    (4,32.62) +- (0,0.55)
    (5,81.79) +- (0,3.14)
    (6,41.10) +- (0,0.80)
    (7,27.98) +- (0,0.29)
    (8,34.97) +- (0,1.00)
    (9,58.09) +- (0,0.60)
};
\addlegendentry{OpenEvolve}
\addplot+[fill=agReq] coordinates {
    (0,70.37) +- (0,0.40)
    (1,31.14) +- (0,0.20)
    (2,97.17) +- (0,0.23)
    (3,112.50) +- (0,0.29)
    (4,31.19) +- (0,0.18)
    (5,79.48) +- (0,2.60)
    (6,45.95) +- (0,2.35)
    (7,28.71) +- (0,1.26)
    (8,34.08) +- (0,0.53)
    (9,64.67) +- (0,0.67)
};
\addlegendentry{White box: key-value trace}
\addplot+[fill=agIO] coordinates {
    (0,68.21) +- (0,1.12)
    (1,31.75) +- (0,0.60)
    (2,98.59) +- (0,0.07)
    (3,114.47) +- (0,0.18)
    (4,28.88) +- (0,0.28)
    (5,80.07) +- (0,4.07)
    (6,40.93) +- (0,1.27)
    (7,27.37) +- (0,1.00)
    (8,33.48) +- (0,0.70)
    (9,62.86) +- (0,0.51)
};
\addlegendentry{White box: I/O traces}
\addplot+[fill=agLRU, bar shift=-8.80pt, forget plot] coordinates {(10,98.96) +- (0,0.70)};
\addplot+[fill=agMG, bar shift=-4.40pt, forget plot] coordinates {(10,99.74) +- (0,0.93)};
\addplot+[fill=agCX, bar shift=0.00pt, forget plot, nodes near coords, point meta=explicit symbolic] coordinates {(10,96.40) +- (0,0.47) [S3FIFO]};
\addplot+[fill=agOE, bar shift=4.40pt, forget plot] coordinates {(10,97.36) +- (0,0.54)};
\addplot+[fill=agIO, bar shift=8.80pt, forget plot] coordinates {(10,98.15) +- (0,0.02)};
\draw[black!40, dashed] (axis cs:9.5,0) -- (axis cs:9.5,123);
\end{axis}
\end{tikzpicture}
  \vspace{-6pt}
  \caption{Average throughput for short runs (45\,s warm-up followed by a 4-minute measurement) on the cache traces. 
  }

  \label{fig:short-agentic-vs-baselines}
  \vspace{-1em}
\end{figure*}
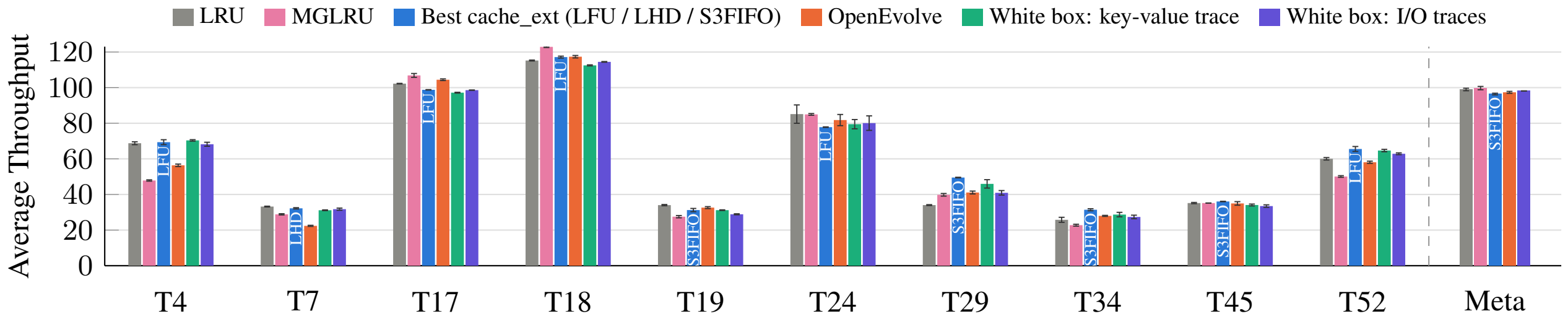

\begin{figure}[t]
  \centering
  \begin{tikzpicture}
  \definecolor{agLRU}{HTML}{8A8A85}%
  \definecolor{agMG}{HTML}{E87BA4}%
  \definecolor{agCX}{HTML}{2A78D6}%
  \definecolor{agOE}{HTML}{EB6834}%
  \definecolor{agReq}{HTML}{1BAF7A}%
  \definecolor{agIO}{HTML}{6250D6}%
\begin{axis}[
    width=\columnwidth, height=3.36cm,
    ybar, bar width=2.9pt,
    xticklabel style={font=\scriptsize},
    ybar=0.5pt,
    ymin=0, ymax=123, ytick distance=20,
    xmin=-0.5, xmax=9.5,
    xtick={0,...,9},
    xticklabels={T4,T7,T17,T18,T19,T24,T29,T34,T45,T52},
    xtick style={draw=none},
    tick label style={font=\footnotesize},
    label style={font=\footnotesize},
    ylabel={Average Throughput},
    ylabel shift=-3pt,
    ymajorgrids, grid style={black!12},
    axis x line*=bottom, axis y line*=left,
    every axis plot/.append style={draw=none},
    error bars/y dir=both, error bars/y explicit,
    error bars/error bar style={black!80, line width=0.35pt},
    error bars/error mark options={rotate=90, mark size=0.9pt, black!80, line width=0.35pt},
    every node near coord/.append style={
        rotate=90, anchor=east, font=\fontsize{5}{6}\selectfont, text=white, inner sep=0.8pt},
    legend style={draw=none, fill=none, font=\fontsize{6}{7}\selectfont, at={(0.46,1.02)}, anchor=south,
                  legend columns=2, legend cell align=left,
                  /tikz/every even column/.append style={column sep=4pt}},
    legend image code/.code={\fill[#1] (0cm,-0.08cm) rectangle (0.22cm,0.1cm);},
]
\addplot+[fill=agReq] coordinates {
    (0,70.37) +- (0,0.40)
    (1,31.14) +- (0,0.20)
    (2,97.17) +- (0,0.23)
    (3,112.50) +- (0,0.29)
    (4,31.19) +- (0,0.18)
    (5,79.48) +- (0,2.60)
    (6,45.95) +- (0,2.35)
    (7,28.71) +- (0,1.26)
    (8,34.08) +- (0,0.53)
    (9,64.67) +- (0,0.67)
};
\addlegendentry{White box: k-v trace, 4 min interval}
\addplot+[fill=agReq!40] coordinates {
    (0,69.05) +- (0,0.60)
    (1,31.01) +- (0,0.09)
    (2,96.12) +- (0,0.14)
    (3,117.17) +- (0,0.51)
    (4,34.02) +- (0,0.60)
    (5,82.69) +- (0,1.44)
    (6,40.01) +- (0,0.32)
    (7,27.21) +- (0,0.71)
    (8,36.52) +- (0,0.33)
    (9,60.22) +- (0,2.28)
};
\addlegendentry{White box: k-v trace, 4\,$\times$\,1 min intervals}
\addplot+[fill=agIO] coordinates {
    (0,68.21) +- (0,1.12)
    (1,31.75) +- (0,0.60)
    (2,98.59) +- (0,0.07)
    (3,114.47) +- (0,0.18)
    (4,28.88) +- (0,0.28)
    (5,80.07) +- (0,4.07)
    (6,40.93) +- (0,1.27)
    (7,27.37) +- (0,1.00)
    (8,33.48) +- (0,0.70)
    (9,62.86) +- (0,0.51)
};
\addlegendentry{White box: I/O traces, 4 min interval}
\addplot+[fill=agIO!40] coordinates {
    (0,67.64) +- (0,0.27)
    (1,27.07) +- (0,0.13)
    (2,96.22) +- (0,0.15)
    (3,112.32) +- (0,0.43)
    (4,28.93) +- (0,0.12)
    (5,82.85) +- (0,1.01)
    (6,37.89) +- (0,0.80)
    (7,27.56) +- (0,0.39)
    (8,33.14) +- (0,0.33)
    (9,63.81) +- (0,1.12)
};
\addlegendentry{White box: I/O traces, 4\,$\times$\,1 min intervals}
\end{axis}
\end{tikzpicture}
  \vspace{-14pt}
  \caption{White box offline approach when running on four 1-minute offline intervals versus one 4-minute interval.} %
  \label{fig:short-agentic-interval}
  \vspace{-1em}
\end{figure}
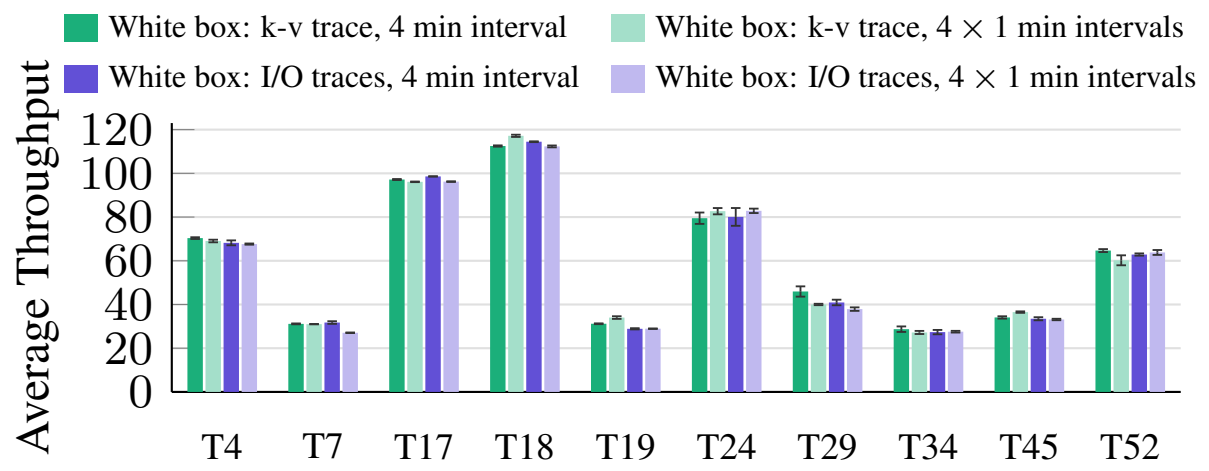

We observe that a significant disadvantage of OpenEvolve-style approaches is that they treat the entire system as a black box and try to optimize the score by tinkering with it. In our case, the black box is quite complex: it is a live server running an operating system, running production traces on a relatively complicated storage system (LevelDB), which may interact with each layer of the system unpredictably.

Therefore, we seek a more ``white box'' form of iterative learning, where we can pass additional information gleaned about the workload or from the previous run to the next iteration besides just the final score. Since OpenEvolve's framework currently prevents adding such feedback loops, we simply ran the same iterative process using Claude Code (Opus 5). In addition, we test how well we can generate policies by augmenting the feedback loop depicted in Figure~\ref{fig:offline_search_loop} with two types of information: the logical key-value accesses (e.g., GET Key=1, SET Key=3 Value=4) of the offline trace, and the OS-based I/O trace collected from the cgroup, specifically the number of read and written bytes, and the page accesses and faults. We instruct the agent to go through a loop of generating a policy, compiling it, loading it, and testing it on a particular offline trace, resulting in an average throughput score. We also ask it to compare the policy to the existing policies and other policies it generated, learn whether it can be improved, and to find a policy that would yield the highest throughput while remaining generalizable (to avoid overfitting).

\paragraph{White box vs. black box.} We plot the average throughput over the 4 minute offline trace in Figure~\ref{fig:short-agentic-vs-baselines}. We compare the ``white box'' approach to OpenEvolve using two types of information: either feeding the model the request trace, or feeding it I/O traces. For the Meta trace, we only developed an I/O trace-based policy, as the trace runs on the filesystem, not a key-value store. The results show that the white box approach on average performs slightly better than OpenEvolve, and the differences are most pronounced in Twitter clusters T4 and T17, where OpenEvolve had significant throughput drops compared to the best \cacheext policy. In addition, we can see there is almost no difference between feeding the model the key-value trace and providing it with raw I/O traces. 

\paragraph{Using different offline intervals.}
Our experiments so far all used a single 4-minute offline trace. In order to confirm that we are not overfitting the caching policies to that particular segment, we run another experiment where we split up the offline trace into four different 1-minute intervals that are taken from different positions in the trace, at 25M, 50M, 75M and 100M requests. The results, presented in Figure~\ref{fig:short-agentic-interval}), yield almost identical performance against the offline traces.

\paragraph{General limitations of offline iterative approaches.} Beyond failing to achieve consistently higher throughput using the offline traces, there are two additional problems with both offline iterative approaches (OpenEvolve and our white box iterative method). First, the approaches are very expensive. Unsurprisingly, hundreds of iterations on a 4 minute trace, where each iteration requires generating eBPF code, compiling it successfully, and loading it, can be quite expensive. We present a full cost analysis in Section \ref{sec:cost}.
Second, the LLM is trying to optimize its score using an offline trace. However, access patterns and workloads can change over time. In the evaluation we show that when run on ``unseen'' online traces, these policies often fail to generalize (\S\ref{sec:eval-offline-unseen}).

\subsection{Limitations of Online Knob Tuning}
\label{sec:knob-tuning}

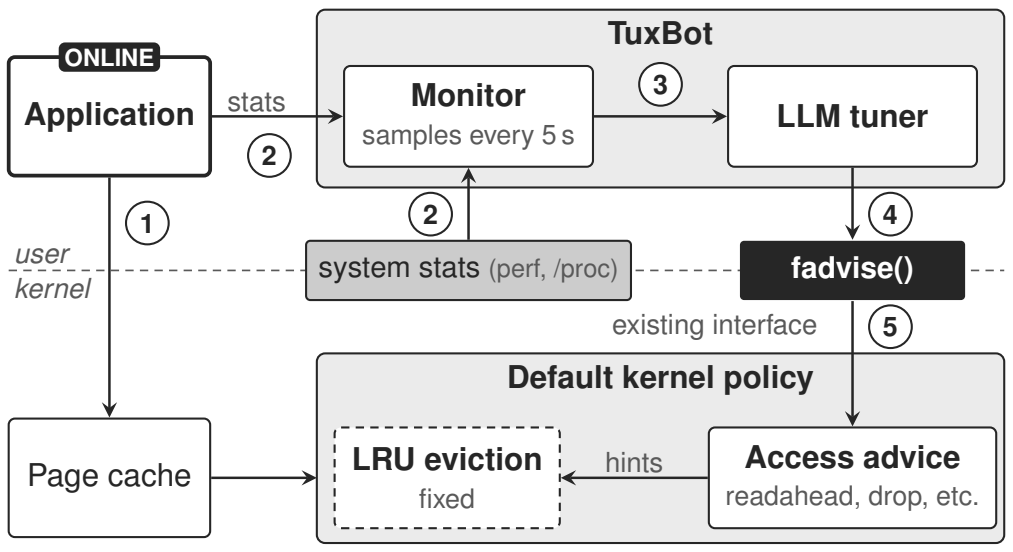
\begin{figure}[t]
  \centering
  \resizebox{0.8\columnwidth}{!}{\begingroup%
\definecolor{archfill}{gray}{0.92}%
\definecolor{archmap}{gray}{0.80}%
\definecolor{archline}{gray}{0.15}%
\definecolor{archmuted}{gray}{0.35}%
\def\fs#1{\fontfamily{phv}\fontsize{\dimexpr #1pt*92/100\relax}%
  {\dimexpr #1pt*110/100\relax}\selectfont}%
\def\note#1{{\fs{7}\color{archmuted}#1}}%
\fontfamily{phv}\selectfont%
\resizebox{\linewidth}{!}{%
\begin{tikzpicture}[
  archline, line width=0.5pt, >=stealth,
  container/.style={draw=archline, fill=archfill, rounded corners=2pt},
  comp/.style={draw=archline, fill=white, rounded corners=1.5pt, align=center,
               inner sep=1.5pt, font=\fs{8.2}},
  map/.style={draw=archline, fill=archmap, rounded corners=1pt, align=center,
              inner sep=1pt, minimum height=0.5cm, minimum width=1.55cm, font=\fs{7.6}},
  iface/.style={draw=none, rounded corners=1pt, align=center, inner sep=2pt,
                minimum height=0.5cm, minimum width=1.9cm, text=white,
                font=\fs{7.8}\bfseries},
  pill/.style={rounded corners=1.5pt, inner sep=1.4pt, fill=archline, text=white,
               font=\fs{6.2}\bfseries},
  sub/.style={font=\fs{7.2}, text=archmuted},
  stepno/.style={circle, draw=archline, fill=white, inner sep=0.4pt,
               minimum size=0.36cm, font=\fs{7}\bfseries},
  flow/.style={->, line width=0.6pt},
]

\draw[archmuted, dash pattern=on 2pt off 1.5pt, line width=0.4pt] (0,2.65) -- (8.4,2.65);
\node[anchor=south west, sub, font=\fs{7.2}\itshape, inner sep=1pt] at (0,2.67) {user};
\node[anchor=north west, sub, font=\fs{7.2}\itshape, inner sep=1pt] at (0,2.63) {kernel};

\node[comp, minimum width=1.7cm, minimum height=1.0cm, line width=0.9pt]
  (in) at (0.85,3.95) {\textbf{Application}};
\node[pill] at (0.85,4.45) {ONLINE};

\draw[container] (2.6,3.35) rectangle (8.4,4.85);
\node[anchor=north, font=\fs{8.2}\bfseries, inner sep=0pt] at (5.5,4.76) {TuxBot};
\node[comp, minimum width=2.1cm, minimum height=0.85cm] (mon) at (3.88,3.95)
  {\textbf{Monitor}\\\note{samples every 5\,s}};
\node[comp, minimum width=2.1cm, minimum height=0.85cm] (tun) at (7.12,3.95)
  {\textbf{LLM tuner}};
\draw[flow] (mon) -- (tun);

\node[map, inner xsep=3pt] (stats) at (3.88,2.65) {system stats {\fs{6.6}\color{archmuted}(perf, /proc)}};
\node[iface, fill=archline] (if) at (7.12,2.65) {fadvise()};
\node[sub, anchor=east] at (6.95,2.175) {existing interface};

\node[comp, minimum width=1.7cm, minimum height=1.0cm] (pc) at (0.85,0.9) {Page cache};
\draw[container] (2.6,0.35) rectangle (8.4,1.95);
\node[anchor=north, font=\fs{8.2}\bfseries, inner sep=0pt] at (5.5,1.86)
  {Default kernel policy};
\node[comp, minimum width=1.9cm, minimum height=0.85cm, dash pattern=on 2pt off 1.5pt]
  (lru) at (3.7,0.9) {\textbf{LRU eviction}\\\note{fixed}};
\node[comp, minimum width=1.9cm, minimum height=0.85cm, inner xsep=4pt]
  (adv) at (7.12,0.9) {\textbf{Access advice}\\\note{readahead, drop, etc.}};

\draw[flow] (in) -- (pc);
\draw[flow] (1.7,3.95) -- node[above, sub, inner sep=1pt, pos=0.35] {stats} (mon.west |- 0,3.95);
\draw[flow] (pc) -- (2.6,0.9);
\draw[flow] (stats) -- (mon);
\draw[flow] (tun) -- (if);
\draw[flow] (if) -- (adv);
\draw[flow] (adv) -- node[above, sub, inner sep=1pt] {hints} (lru);

\node[stepno] at (1.17,3.05) {1};
\node[stepno] at (2.2,3.62) {2};
\node[stepno] at (3.56,3.125) {2};
\node[stepno] at (5.5,4.22) {3};
\node[stepno] at (7.44,3.125) {4};
\node[stepno] at (7.44,2.175) {5};

\end{tikzpicture}}%
\endgroup%
}
  \caption{TuxBot tuning fadvise() online. As the application runs~\textcircled{\scriptsize 1}, TuxBot observes stats~\textcircled{\scriptsize 2}, picks a value~\textcircled{\scriptsize 3} and sets it via \texttt{fadvise()}~\textcircled{\scriptsize 4}, providing hints to the fixed LRU policy~\textcircled{\scriptsize 5}.}
  \label{fig:tuxbot_fadvise}
  \vspace{-1em}
\end{figure}

A natural alternative to writing new policies is to tune the knobs the kernel
already exposes while the workload runs.
TuxBot~\citep{liargkovas2026tuxbot} does this with an LLM in the loop. However, Tuxbot is currently limited to a certain set of OS parameters and interfaces, namely the default scheduler, power, and writeback knobs, which do not control the page cache.
We therefore augmented Tuxbot and gave it access to the interface that most directly controls the page cache during application runtime, the \texttt{posix\_fadvise} access hint, applied live to every data file, and let it tune the hint with throughput feedback~(Figure~\ref{fig:tuxbot_fadvise}).

At first glance the result seems impressive, as TuxBot raised throughput on T24 by up to 77\%.
On closer inspection, every run converged to the same hint,
\texttt{FADV\_RANDOM}, which simply disables readahead. While disabling readahead can be advantageous to the traces we measured (traces that exhibit significant random read access patterns), it could significantly harm performance for scan-heavy workloads.
Interestingly, none of the other five hints helped, and with the random hint removed TuxBot gained only 3\%. Online tuning added nothing beyond that single setting.
Setting \texttt{FADV\_RANDOM} by default did as well or better, because TuxBot lost time trying other hints before settling on it.

The gain thus comes from switching off one mechanism, not from tuning.
Readahead fills a small cache with pages that are never read, so disabling it
helps by 37--63\% on the Twitter clusters where this happens (T24, T34, T52),
but by at most 2\% on T17 and Meta.
On workloads where readahead is not the problem, the hint has nothing to offer,
yet the choice of eviction policy still matters.
On T17, for example, no hint improves throughput by more than
2\%, whereas switching to MGLRU improves it by 9\%. Disabling readahead can
even backfire, lowering MGLRU's throughput by 9--13\% on T17 and T18.
Knobs such as \texttt{fadvise()} make coarse-grained decisions, but they cannot choose which pages stay in the cache.
That choice belongs to the eviction policy, which is what \cacheext lets us
customize. In the rest of the paper we treat readahead as orthogonal.
\section{Design: Online Learning with \sys}
\label{sec:design}

\begin{figure}[t]
  \centering
  \resizebox{0.8\columnwidth}{!}{\begingroup%
\definecolor{archfill}{gray}{0.92}%
\definecolor{archmap}{gray}{0.80}%
\definecolor{archline}{gray}{0.15}%
\definecolor{archmuted}{gray}{0.35}%
\def\fs#1{\fontfamily{phv}\fontsize{\dimexpr #1pt*92/100\relax}%
  {\dimexpr #1pt*110/100\relax}\selectfont}%
\def\note#1{{\fs{7}\color{archmuted}#1}}%
\def\symge{{\fontencoding{U}\fontfamily{psy}\selectfont\char179}}%
\input{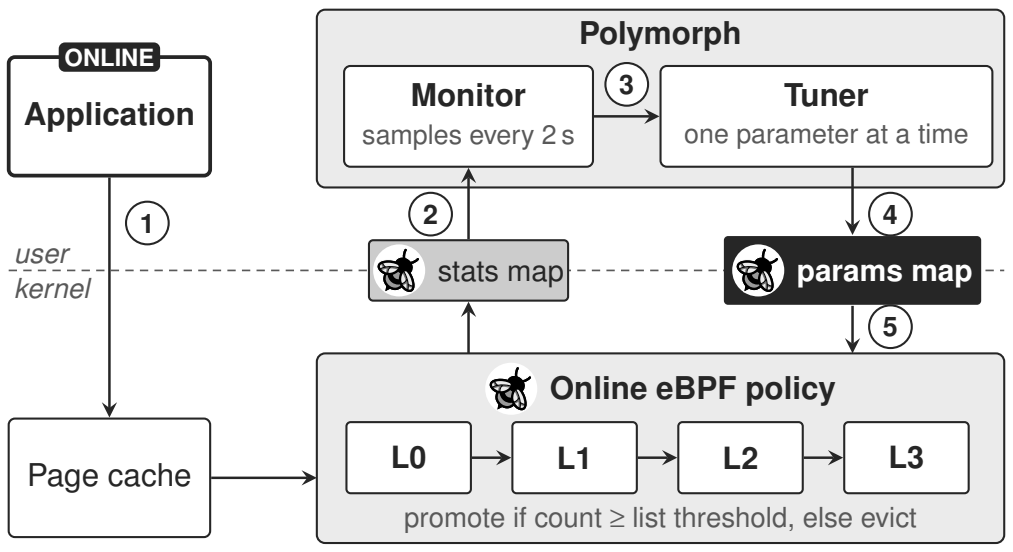}%
\fontfamily{phv}\selectfont%
\resizebox{\linewidth}{!}{%
\begin{tikzpicture}[
  archline, line width=0.5pt, >=stealth,
  container/.style={draw=archline, fill=archfill, rounded corners=2pt},
  comp/.style={draw=archline, fill=white, rounded corners=1.5pt, align=center,
               inner sep=1.5pt, font=\fs{8.2}},
  map/.style={draw=archline, fill=archmap, rounded corners=1pt, align=center,
              inner sep=1pt, minimum height=0.5cm, minimum width=1.55cm, font=\fs{7.6}},
  iface/.style={draw=none, fill=archline, rounded corners=1pt, align=center,
                inner sep=2pt, minimum height=0.5cm, minimum width=1.9cm, text=white,
                font=\fs{7.8}\bfseries},
  lst/.style={draw=archline, fill=white, rounded corners=1pt, minimum width=1.05cm,
              minimum height=0.6cm, font=\fs{8}\bfseries},
  pill/.style={rounded corners=1.5pt, inner sep=1.4pt, fill=archline, text=white,
               font=\fs{6.2}\bfseries},
  sub/.style={font=\fs{7.2}, text=archmuted},
  stepno/.style={circle, draw=archline, fill=white, inner sep=0.4pt,
               minimum size=0.36cm, font=\fs{7}\bfseries},
  flow/.style={->, line width=0.6pt},
]

\draw[archmuted, dash pattern=on 2pt off 1.5pt, line width=0.4pt] (0,2.65) -- (8.4,2.65);
\node[anchor=south west, sub, font=\fs{7.2}\itshape, inner sep=1pt] at (0,2.67) {user};
\node[anchor=north west, sub, font=\fs{7.2}\itshape, inner sep=1pt] at (0,2.63) {kernel};

\node[comp, minimum width=1.7cm, minimum height=1.0cm, line width=0.9pt]
  (in) at (0.85,3.95) {\textbf{Application}};
\node[pill] at (0.85,4.45) {ONLINE};

\draw[container] (2.6,3.35) rectangle (8.4,4.85);
\node[anchor=north, font=\fs{8.2}\bfseries, inner sep=0pt] at (5.5,4.76) {\sys};
\node[comp, minimum width=2.1cm, minimum height=0.85cm] (mon) at (3.88,3.95)
  {\textbf{Monitor}\\\note{samples every 2\,s}};
\node[comp, minimum width=2.8cm, minimum height=0.85cm] (tun) at (6.9,3.95)
  {\textbf{Tuner}\\\note{one parameter at a time}};
\draw[flow] (mon) -- (tun);

\node[map] (stats) at (3.88,2.65) {\raisebox{\dimexpr -0.5\height+2.5pt\relax}{\ebpfbee}\hspace{3pt}stats map};
\node[iface] (params) at (7.12,2.65)
  {\raisebox{\dimexpr -0.5\height+2.5pt\relax}{\ebpfbee}\hspace{3pt}params map};

\node[comp, minimum width=1.7cm, minimum height=1.0cm] (pc) at (0.85,0.9) {Page cache};
\draw[container] (2.6,0.35) rectangle (8.4,1.95);
\node[anchor=north, font=\fs{8.2}\bfseries, inner sep=0pt] at (5.5,1.88)
  {\raisebox{\dimexpr -0.5\height+2.5pt\relax}{\ebpfbee}\hspace{3pt}Online eBPF policy};
\foreach \nm [count=\i from 0] in {L0, L1, L2, L3}
  \node[lst] (L\i) at ({3.375 + 1.4*\i}, 1.07) {\nm};
\foreach \a/\b in {0/1, 1/2, 2/3} \draw[flow] (L\a) -- (L\b);
\node[sub] at (5.5,0.55)
  {promote if count \symge\ list threshold, else evict};

\draw[flow] (in) -- (pc);
\draw[flow] (pc) -- (2.6,0.9);
\draw[flow] (3.88,1.95) -- (stats);
\draw[flow] (stats) -- (mon);
\draw[flow] (tun.south -| params) -- (params);
\draw[flow] (params) -- (7.12,1.95);

\node[stepno] at (1.17,3.05) {1};
\node[stepno] at (3.56,3.125) {2};
\node[stepno] at (5.215,4.22) {3};
\node[stepno] at (7.44,3.125) {4};
\node[stepno] at (7.44,2.175) {5};

\end{tikzpicture}}%
\endgroup%
}
  \vspace{-1em}
  \caption{\sys's online policy. I/O~\textcircled{\scriptsize 1} reaches the eBPF policy~\textcircled{\scriptsize 2}. The controller periodically reads stats~\textcircled{\scriptsize 3} and re-tunes a parameter~\textcircled{\scriptsize 4}\textcircled{\scriptsize 5}.}
  \label{fig:online_policy_arch_min}
  \vspace{-1em}
\end{figure}

\textbf{Overview.} We introduce \sys, a novel approach to automating OS policies that can adapt online to live workloads. \sys applies the same design to the page cache and to the process scheduler. For each policy, it consists of a generalized kernel policy, loaded through eBPF, whose behavior is set by parameters in a map shared with user space, and a user mode component, the \textit{tuner}, which reads system statistics every 2 seconds and writes new parameter values into the map (Figure~\ref{fig:online_policy_arch_min}). The kernel policy reads the map at its next decision, so it adapts continuously to the running workload without being reloaded. In the offline loop of Figure~\ref{fig:offline_search_loop}, an LLM generates and evaluates a new version of the policy in every iteration. In \sys, an agent writes the policy and the tuner once, and importantly, \emph{no LLM is called during the online loop}. The two policies differ in what they control and in what their tuners read. The cache policy, built on \cacheext, manages a hierarchy of four lists, and its tuner reads I/O, page, and cgroup statistics and adjusts one parameter at a time. The scheduler policy, built on sched\_ext, manages run queues, wakeup priority, and time slices, and its tuner reads CPU utilization and CPU time per wakeup and selects one of three settings (\S\ref{sec:design-sched}).

Figure~\ref{fig:online_policy_arch_min} shows the main components and the flow over one iteration for the page cache. The kernel policy is loaded into the kernel through eBPF and \cacheext, and it manages the pages of the running workload in a hierarchy of four lists~\textcircled{\scriptsize 1}. The user mode tuner reads I/O, page, and cgroup statistics~\textcircled{\scriptsize 2}, selects a parameter to adjust in sequential order~\textcircled{\scriptsize 3}, and updates the shared parameter map~\textcircled{\scriptsize 4}. The new values apply to the kernel policy~\textcircled{\scriptsize 5}. In the next iteration, the tuner checks the result and decides whether to keep the adjustment or revert it.

\subsection{Generalized Cache Policy}

We seek OS policies that can be adapted to each workload. However, as we show in the next section, the offline approaches do not hold up on unseen data. Learning must therefore be continuous and online, while the workload runs. The challenge is that we require very lightweight mechanisms (in terms of cost and performance overhead) to switch between different policies in response to workload changes. 

Our main observation is that across the existing cache policies and the ones generated by offline learning, the variation between policies can be described by a small set of parameters: the number of lists, the rules for promoting or evicting folios between them, and what to do on a folio access. 
This led us to design a general cache data structure that can represent a wide variety of caching policies by changing only its parameters, without reloading or flushing the policy as it adapts to workload changes. 

The policy is a hierarchy of four lists, and it promotes or evicts folios between the lists. Four parameters are adjustable: the access-count threshold for promotion or eviction in each list, the weight, or ratio of sizes, between the lists, the initial access count of a folio in each list, and the frequency cap at which each list stops promoting. We chose these parameters to make the policy as granular as possible, so that selecting the right parameter values yields the right policy. ``Classic'' caching policies can easily be implemented by this structure: for example, segmented LRU (SLRU)~\cite{slru} can be approximated by treating each list as a separate LRU queue, and the movement between the queues and eviction can be controlled using the lists' access counts.

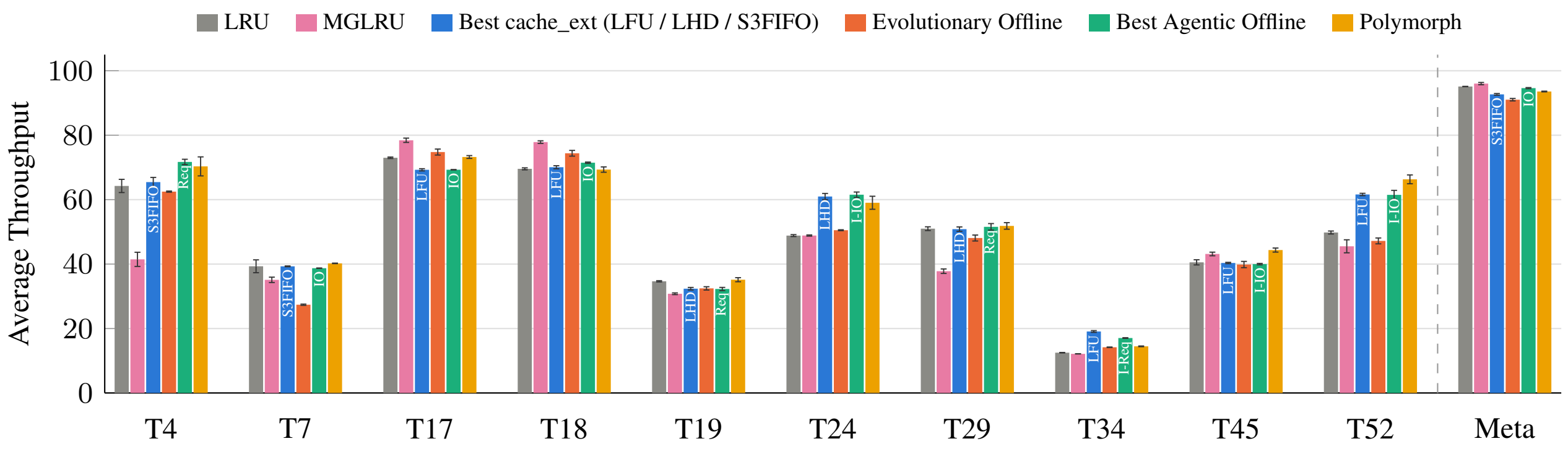
\begin{figure*}[t]
  \centering
  \begin{tikzpicture}
\begin{axis}[
    width=\textwidth, height=5.2cm,
    ybar=0.5pt, bar width=4.3pt,
    ymin=0, ymax=105, ytick distance=20,
    xmin=-0.42, xmax=10.42,
    xtick={0,...,10},
    xticklabels={T4,T7,T17,T18,T19,T24,T29,T34,T45,T52,Meta},
    xtick style={draw=none},
    tick label style={font=\footnotesize},
    label style={font=\footnotesize},
    ylabel={Average Throughput},
    ylabel shift=-3pt,
    ymajorgrids, grid style={black!12},
    axis x line*=bottom, axis y line*=left,
    every axis plot/.append style={draw=none},
    error bars/y dir=both, error bars/y explicit,
    error bars/error bar style={black!80, line width=0.35pt},
    error bars/error mark options={rotate=90, mark size=0.9pt, black!80, line width=0.35pt},
    every node near coord/.append style={
        rotate=90, anchor=east, font=\fontsize{4.5}{5}\selectfont, text=white, inner sep=0.8pt},
    legend style={draw=none, fill=none, font=\scriptsize, at={(0.5,1.02)}, anchor=south,
                  legend columns=6, legend cell align=left,
                  /tikz/every even column/.append style={column sep=6pt}},
    legend image code/.code={\fill[#1] (0cm,-0.08cm) rectangle (0.22cm,0.1cm);},
]
\addplot+[fill=evLRU] coordinates {
    (0,64.25) +- (0,2.05)
    (1,39.32) +- (0,1.99)
    (2,72.99) +- (0,0.23)
    (3,69.56) +- (0,0.31)
    (4,34.66) +- (0,0.19)
    (5,48.84) +- (0,0.30)
    (6,50.99) +- (0,0.58)
    (7,12.50) +- (0,0.07)
    (8,40.54) +- (0,0.80)
    (9,49.80) +- (0,0.46)
    (10,95.11) +- (0,0.05)
};
\addlegendentry{LRU}
\addplot+[fill=evMG] coordinates {
    (0,41.47) +- (0,2.20)
    (1,35.11) +- (0,0.83)
    (2,78.42) +- (0,0.67)
    (3,77.86) +- (0,0.40)
    (4,30.79) +- (0,0.27)
    (5,48.86) +- (0,0.18)
    (6,37.79) +- (0,0.70)
    (7,12.14) +- (0,0.04)
    (8,43.14) +- (0,0.56)
    (9,45.52) +- (0,2.02)
    (10,96.01) +- (0,0.35)
};
\addlegendentry{MGLRU}
\addplot+[fill=evCX, nodes near coords, point meta=explicit symbolic] coordinates {
    (0,65.44) +- (0,1.46) [S3FIFO]
    (1,39.30) +- (0,0.12) [S3FIFO]
    (2,69.26) +- (0,0.34) [LFU]
    (3,70.06) +- (0,0.47) [LFU]
    (4,32.36) +- (0,0.39) [LHD]
    (5,60.97) +- (0,0.95) [LHD]
    (6,50.84) +- (0,0.69) [LHD]
    (7,19.08) +- (0,0.27) [LFU]
    (8,40.34) +- (0,0.20) [LFU]
    (9,61.60) +- (0,0.36) [LFU]
    (10,92.67) +- (0,0.26) [S3FIFO]
};
\addlegendentry{Best \cacheext (LFU / LHD / S3FIFO)}
\addplot+[fill=evOE] coordinates {
    (0,62.49) +- (0,0.17)
    (1,27.39) +- (0,0.19)
    (2,74.77) +- (0,0.93)
    (3,74.39) +- (0,0.89)
    (4,32.46) +- (0,0.51)
    (5,50.51) +- (0,0.14)
    (6,48.09) +- (0,0.92)
    (7,14.18) +- (0,0.08)
    (8,39.86) +- (0,0.97)
    (9,47.19) +- (0,0.91)
    (10,91.02) +- (0,0.38)
};
\addlegendentry{Evolutionary Offline}
\addplot+[fill=evOFF, nodes near coords, point meta=explicit symbolic] coordinates {
    (0,71.70) +- (0,0.83) [Req]
    (1,38.72) +- (0,0.07) [IO]
    (2,69.28) +- (0,0.06) [IO]
    (3,71.46) +- (0,0.20) [IO]
    (4,32.27) +- (0,0.49) [Req]
    (5,61.53) +- (0,0.83) [I-IO]
    (6,51.56) +- (0,1.01) [Req]
    (7,17.05) +- (0,0.10) [I-Req]
    (8,39.99) +- (0,0.22) [I-IO]
    (9,61.48) +- (0,1.40) [I-IO]
    (10,94.61) +- (0,0.17) [IO]
};
\addlegendentry{Best Agentic Offline}
\addplot+[fill=evON] coordinates {
    (0,70.32) +- (0,2.95)
    (1,40.22) +- (0,0.08)
    (2,73.23) +- (0,0.44)
    (3,69.33) +- (0,0.83)
    (4,35.15) +- (0,0.63)
    (5,59.03) +- (0,2.02)
    (6,51.84) +- (0,1.02)
    (7,14.47) +- (0,0.11)
    (8,44.35) +- (0,0.63)
    (9,66.30) +- (0,1.37)
    (10,93.56) +- (0,0.11)
};
\addlegendentry{\sys}
\draw[black!40, dashed] (axis cs:9.5,0) -- (axis cs:9.5,105);
\end{axis}
\end{tikzpicture}
  \vspace{-6pt}
  \caption{Average throughput on the 11 cache traces.
  For \cacheext and Agentic Offline, the bar shows the best-performing variant on that workload, and its name is printed inside the bar (I-Req and I-IO are the interval request-data and interval I/O-signal variants).}
  \label{fig:eval-throughput-by-cluster}
\end{figure*}

\subsection{User Mode Tuner}
While the kernel policy is loaded and the workload runs, the user mode tuner selects adjustments that customize the policy to the changing patterns of the workload, and it applies them by updating the parameters through a map shared between the kernel and user space. The generalized policy therefore stays in place, and each adjustment adds minimal overhead. The tuner reads the signals that are relevant and available from user mode: per-CPU counters and list sizes, the I/O statistics of the block device, and the memory statistics of the cgroup.

The tuner then scores the current configuration by page touches per second. Then it chooses to tune one of the available parameters at a time. A trial change is kept only if it beats a freshly measured baseline by more than 4\% twice in a row; otherwise, it is reverted and the search direction flips. After a second rejection, the tuner holds the current setting, and if performance falls below 95\% of the best seen, it restores the best configuration. Communication flows through shared BPF maps: statistics flow to the user mode tuner, which writes new parameters directly into the memory-mapped parameter struct of the policy, and the new values take effect on the next eviction scan, with no reload or re-attach.
This logic lets the tuner respond to changes in the workload's patterns continuously, without disrupting performance as long as the changes do not show clear gains.

Crucially, \sys never calls LLMs during its runtime, and all of its decisions come from the dynamic policy that is controlled by the tuner, deterministically. To set the initial parameter values of the kernel policy and the user mode tuner, we used an AI agent (Claude) for a one-shot analysis of one cluster, T34, which we chose for its complexity. In the next section, we show that even though we tuned the initial parameters on only one cluster, the throughput gains hold across multiple clusters.

\subsection{Generalized Scheduler Policy}
\label{sec:design-sched}
We now show how \sys can be applied to the scheduler use case. We observe that the scheduling policies in Table~\ref{tab:sched_motivation} differ in three decisions: whether all CPUs share one run queue or each socket has its own, how far a task that has just woken up may move ahead of tasks that are already waiting, and how long a task runs before it returns to the queue. 

Therefore, \sys's scheduler is one sched\_ext policy that keeps these three decisions as parameters in a shared eBPF map (Appendix~\ref{app:sched}): the queue scope (shared or per socket), the wake boost (how much waiting time a woken task may skip, from 0 to 1\,s), and the time slice (from 0.05 to 20\,ms). Every scheduling callback reads the map, so a new setting applies from the next scheduling decision, without reloading the policy. Long-slice, wake-boost, and per-socket are settings of this one policy.

The tuner reads only system counters, never the application's latency or throughput (since these are not often not reported by the application). Every 2 seconds, it reads three perf counters on the workload's CPUs and computes two statistics: the number of busy CPUs, and the CPU time per wakeup, which is the busy CPU time divided by the number of task wakeups. When the CPUs are at least 99\% busy and each wakeup brings 400\,$\mu$s of CPU time or more, the workload is CPU-bound, and the tuner selects Long-slice. When the CPUs are at least 99\% busy with shorter runs per wakeup, request threads compete with batch work, and it selects Wake-boost. When the CPUs are less busy, it selects Per-socket. Similar to the caching use case, an AI coding agent (Claude Code) wrote the policy and the tuner once, from runs of the four workloads in Table~\ref{tab:sched_motivation}, and the same code serves every workload. Appendix~\ref{app:sched} lists the counters and thresholds.

\subsection{Other Subsystems}
\label{sec:design-other}
The same loop applies to other OS policies~\cite{qin2026policies} when three conditions hold. First, userspace must be able to replace the policy at runtime, through an eBPF extension point, as for prefetching~\cite{fetchbpf}, page fault handling~\cite{bpf_fault}, and in-kernel storage functions~\cite{xrp}, or through a framework that delegates the policy to user space~\cite{ghost,syrup}. Second, the policy must read its parameters from memory it shares with the tuner (eBPF map), so that they can change without a reload. Third, the tuner needs system statistics that reflect how the workload uses the subsystem, such as hardware counters or tracepoints.

\section{Evaluation}
\label{sec:evaluation}

We evaluate how \sys compares with existing approaches for generating OS policies with AI, in performance and in cost, first for the page cache and then for the process scheduler (\S\ref{sec:eval-sched}). We answer three questions. \textbf{Q1:} How does \sys perform compared with existing caching policies and with policies that AI generates offline? \textbf{Q2:} Does \sys generalize to unseen online workloads? \textbf{Q3:} How does the cost of \sys compare with existing AI-based approaches?

\subsection{Case Study 1: Page Caching}

\textbf{System configuration}. We run our page cache experiments on Cloudlab~\cite{cloudlab} c6525-25g machines, each with a 16-core AMD Rome CPU, 128GB of memory, and a 480GB SSD, running Ubuntu 22.04 with Linux v6.6.8. To make our results reproducible, we pin each experiment to its CPUs, disable SMT, swap, and address space randomization, and drop the page cache before each test.

\textbf{Workloads and baselines}. We use 10 clusters of the Twitter cache workloads~\cite{258935} and the Meta block storage workload~\cite{MetaPlatforms2026}, which are real-world production records of cache accesses with varied types, patterns, and scale. Each cluster runs on LevelDB with a memory limit of 10\% of the cluster's data size, so that the page cache evicts pages regularly. We measure average throughput and compare against the kernel's default LRU policy, the kernel's MGLRU, and the LHD, LFU, and S3FIFO policies implemented on \cacheext.

\textbf{Test runs}. A \emph{short} run has a 45-second warm-up followed by a 4-minute measurement. A \emph{long} run has a 5-minute warm-up followed by a 25-minute measurement, on a part of the trace that starts after the data of the short run and does not overlap it.

\begin{figure}[t]
  \centering
  \begin{tikzpicture}
\begin{axis}[
    scale only axis, width=6.95cm, height=2.2cm,
    ybar, bar width=9pt, bar shift=0pt,
    ymin=-8, ymax=8.5, ytick={-8,-4,0,4,8},
    xmin=-0.55, xmax=9.55,
    xtick={0,...,9},
    xticklabels={{MGLRU},{S3-\\FIFO},{LFU},{LHD},{Evol.\\Offline},{Off-\\Req$^\dagger$},{Off-\\IO},{Off-\\IReq$^\dagger$},{Off-\\IIO$^\dagger$},{\sys}},
    xticklabel style={align=center, text depth=0pt, inner sep=1pt, font=\fontsize{6}{7}\selectfont},
    xtick style={draw=none},
    yticklabel style={font=\scriptsize},
    ylabel={$\Delta$ throughput\\vs.\ LRU (\%)},
    ylabel style={font=\scriptsize, align=center},
    ylabel shift=-4pt,
    ymajorgrids, grid style={black!12},
    axis x line*=bottom, x axis line style={draw=none}, axis y line*=left,
    every axis plot/.append style={draw=none},
    clip=false,
]
\addplot+[fill=evMG] coordinates {(0,-6.17)}; %
\addplot+[fill=evCX] coordinates {(1,1.21)}; %
\addplot+[fill=evCX] coordinates {(2,3.91)}; %
\addplot+[fill=evCX] coordinates {(3,4.24)}; %
\addplot+[fill=evOE] coordinates {(4,-2.74)}; %
\addplot+[fill=evOFF] coordinates {(5,-2.95)}; %
\addplot+[fill=evOFF] coordinates {(6,-0.53)}; %
\addplot+[fill=evOFF] coordinates {(7,2.67)}; %
\addplot+[fill=evOFF] coordinates {(8,2.85)}; %
\addplot+[fill=evON] coordinates {(9,8.39)}; %
\draw[black!80, line width=0.4pt] (axis cs:-0.55,0) -- (axis cs:9.55,0);
\end{axis}
\end{tikzpicture}
  \vspace{-1em}
  \caption{Average throughput difference from the baseline kernel LRU across the 11 workloads. Policies marked $^\dagger$Averaged over the ten Twitter clusters only, because the Meta workload has a single agentic offline (I/O-signal) policy.}
  \label{fig:eval-pct-vs-lru}
  \vspace{-1em}
\end{figure}
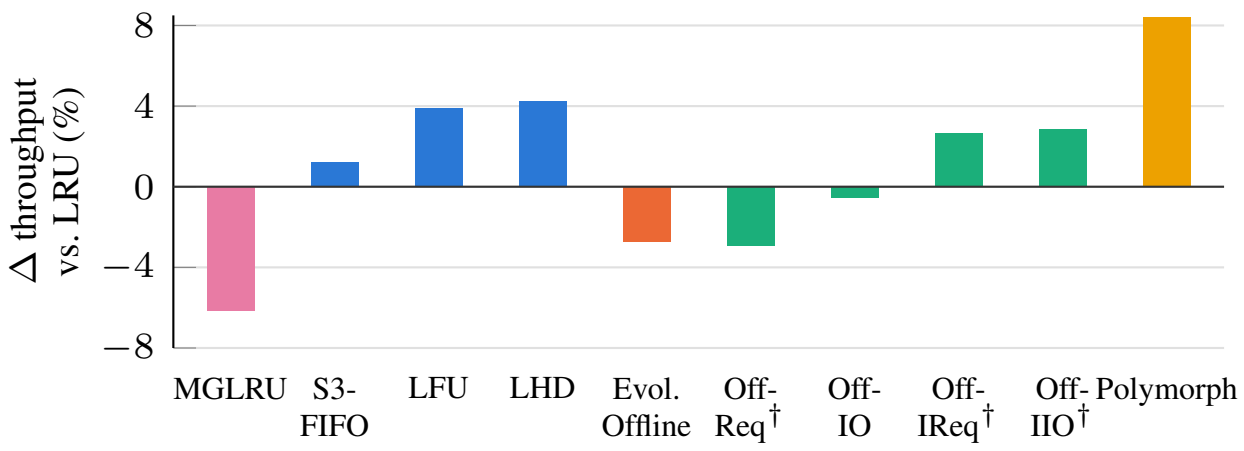

\subsubsection{Offline Search Policies (Q1)}
\label{sec:eval-offline-unseen} 

In \S\ref{subsec:offline_learning} and Figures~\ref{fig:short-agentic-vs-baselines} and~\ref{fig:short-agentic-interval} we showed experiments where offline policy generation tries to outperform existing policies on data it has seen. The offline approaches generated their policies from the short runs, and we test the policies on the same short runs. While the approaches were competitive with existing policies, they did not significantly outperform them, even on the unseen data.

\fbox{%
  \begin{minipage}{\dimexpr\linewidth-2\fboxsep-2\fboxrule\relax}{
      \textbf{Takeaway 1:} LLM-based evolutionary search and AI agents can generate cache policies that are valid eBPF programs that are competitive on real-world workloads.
  } \end{minipage}%
}

\subsubsection{Offline Search on Unseen Workloads (Q2)}
Next, we ask how the offline policies perform on data they have not seen, using the long runs. The results are presented in Figure~\ref{fig:eval-throughput-by-cluster}. The performance of the offline policies drops, and they rarely outperform LRU or the \cacheext policies.
In aggregate (Figure~\ref{fig:eval-pct-vs-lru}), every offline policy trails the static \cacheext policies LFU and LHD. The best offline policy improves throughput by 2.9\% over LRU, compared with 3.9\% for LFU and 4.2\% for LHD. The intervals somewhat prevent overfitting, but the offline policies still do not underperform the \cacheext LFU and LHD policies.

\fbox{%
  \begin{minipage}{\dimexpr\linewidth-2\fboxsep-2\fboxrule\relax}{
      \textbf{Takeaway 2:} The offline approach cannot adequately address unseen data and in the generic case does not create a gain compared to existing approaches.
  } \end{minipage}%
}

\subsubsection{How Does \sys  Perform (Q2)}
\label{sec:eval-online}
On the same long runs, \sys achieves the highest throughput on average, and best or tied on 5 of 11 clusters (Figure~\ref{fig:eval-throughput-by-cluster}). In aggregate, it improves throughput over LRU by 8.4\%, nearly twice the gain of the best other policy, LHD, at 4.2\% (Figure~\ref{fig:eval-pct-vs-lru}). The generalized policy can represent the policy that each workload needs by changing only its parameters, and the user mode tuner learns from I/O signals when and where to adjust the parameters it shares with the kernel policy. As a result, \sys tailors the policy to each workload automatically, without any further calls to an LLM or an agent.

\fbox{%
  \begin{minipage}{\dimexpr\linewidth-2\fboxsep-2\fboxrule\relax}{
      \textbf{Takeaway 3:} \sys does best due to its ability to generically adjust to workloads. %
  } \end{minipage}%
}

\begin{figure*}[t]
  \centering
  \input{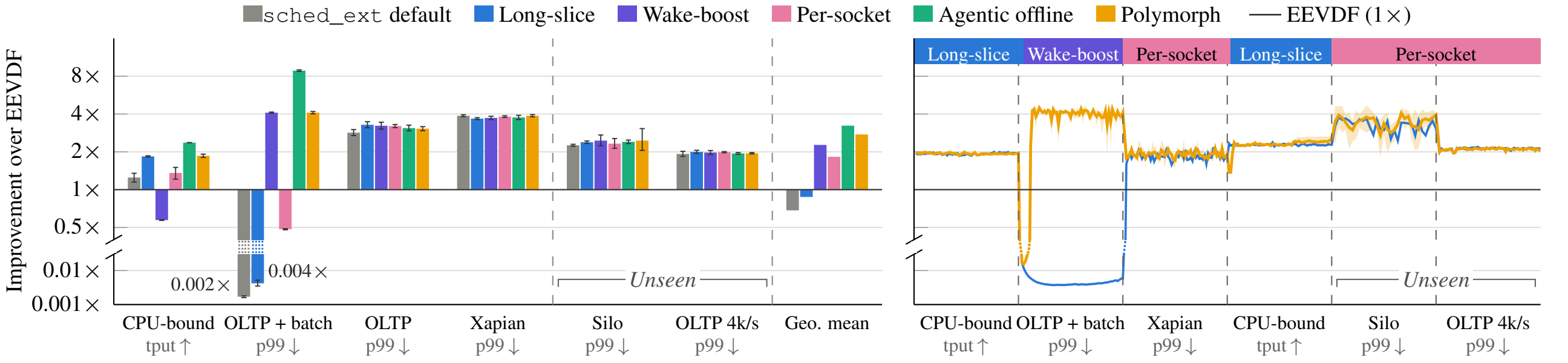}
  \vspace{-8pt}
  \caption{Mean improvement over EEVDF. (a) Per-run end-to-end improvement. (b) Per-second improvement, across 60\,s.
  }
  \label{fig:sched-online}
  \vspace{-1em}
\end{figure*}

\subsubsection{Cost (Q3)}
\label{sec:cost}

\begin{table}[htbp]
  \centering
  \resizebox{\columnwidth}{!}{%
  \begin{tabular}{llrrl}
    \toprule
    Approach & Recurrence & Tokens & Dollars & Duration \\
    \midrule
    OpenEvolve (Offline) & Per Cluster & 10M        & \$300          & 12--24 Hours \\
    White Box (Offline)      & Per Cluster & 125M--240M & \$3750--\$7200 & 1--3 Days    \\
    \sys                 & One time    & 16M        & \$480          & $<$1 Hour    \\
    \bottomrule
  \end{tabular}%
  }
  \vspace{-1em}
  \caption{Cost range comparison of approaches.}
  \label{tab:cache-cost}
  \vspace{-1em}
\end{table}

The approaches widely differ in their cost and the time it takes to generate the policies (Table~\ref{tab:cache-cost}). With OpenEvolve and the white box, each iteration makes one LLM call, which generates a mutation of the code, and then runs the evaluation score function, which does not use AI. Importantly, this search must be repeated for each cluster. The white box approach is much more expensive (and time consuming) than ``vanilla'' open evolve, because it has to incorporate and analyze white box signals from the workload. 

In contrast, \sys needs a single agent prompt, which receives the policy structure and generates the kernel policy and the user mode tuner. The agent also sets the default values of the tuner's hyperparameters in one shot, by reading only the requests of the short run of one cluster (T34, chosen for its complexity). Crucially, the same code then serves every workload, so \sys pays this cost \emph{only once}: amortized over our 11 clusters, it uses about 1.5M tokens per cluster, 7$\times$ fewer than evolutionary search and 86--165$\times$ fewer than agentic search.

\subsection{Case Study 2: Process Scheduler}
\label{sec:eval-sched}

We next evaluate \sys on a second OS policy, the process scheduler, and answer Q1 and Q2 for it.

\textbf{System configuration}. We use a Cloudlab~\cite{cloudlab} c6320 machine (two 14-core Intel Xeon E5-2683 v3 CPUs, 256GB of memory) running Ubuntu 22.04 with Linux v6.17, and pin each workload to 11 CPU threads.

\textbf{Online scheduler}. We evaluate the scheduler of \S\ref{sec:design-sched}, whose tuner switches one \texttt{sched\_ext} policy among the Long-slice, Wake-boost, and Per-socket settings of Table~\ref{tab:sched_motivation}.

\textbf{Offline baseline}. Following the offline loop of Figure~\ref{fig:offline_search_loop}, a separate, sandboxed Claude Code session wrote a \texttt{sched\_ext} policy from scratch, scoring each attempt by its average improvement over EEVDF on the same four workloads (with seeds our test runs do not use); we deploy its best policy unchanged.

We evaluate on the four \emph{seen} workloads of Table~\ref{tab:sched_motivation}, in runs of 15--45 seconds after a short warm-up, and in the same way on two \emph{unseen} workloads that neither agent ran: Silo, an in-memory TPC-C database from TailBench~\cite{kasture2016tailbench}, and the PostgreSQL OLTP workload alone at 4k transactions/s. A \emph{changing} run plays six 60-second stages back to back (CPU-bound, OLTP + batch, Xapian, CPU-bound, Silo, OLTP 4k) without telling the tuner when a stage begins; the last two run separately, from the setting at the end of the fourth. We measure batch throughput on CPU-bound and p99 latency on the others, and compare \sys with EEVDF, the four \texttt{sched\_ext} policies, and the agentic offline policy.

\subsubsection{Offline and Online Search Performance}
On the four seen workloads, \sys matches the best \texttt{sched\_ext} policy on three and comes within 7\% of it on OLTP (Figure~\ref{fig:sched-online}a). The best policy differs between workloads, and each \texttt{sched\_ext} policy is worse than EEVDF on at least one of them (\S\ref{sec:motivation}). The tuner tells these cases apart from its two statistics and selects a suitable setting within 2.5 seconds of the start. In aggregate (Figure~\ref{fig:sched-online}a, right), \sys improves on EEVDF by 2.75$\times$, compared with 2.27$\times$ for the best single policy, Wake-boost, and 0.68$\times$ for \texttt{sched\_ext}'s default policy.

The agentic offline policy reaches 3.23$\times$, mostly on the seen workloads (3.95$\times$ against 3.08$\times$ for \sys): it classifies each task by what woke it, which cuts p99 latency on OLTP + batch to less than half of \sys's, and on CPU-bound it raises batch throughput by nearly stopping the threads that compute briefly and sleep, which the metric does not count. Its search used 3.1M tokens; a fresh agent regenerated \sys's policy and tuner from one prompt with 1.2M.

\fbox{%
  \begin{minipage}{\dimexpr\linewidth-2\fboxsep-2\fboxrule\relax}{
      \textbf{Takeaway 4:} \sys's scheduler adapts one \texttt{sched\_ext} policy to each workload and stays close to the best fixed policy on each. A bespoke offline policy does better on its training workloads, partly by starving unmeasured work.
  } \end{minipage}%
}

\subsubsection{Offline and Online Search with Unseen Workloads}
On the two unseen workloads, \sys also matches the best \texttt{sched\_ext} policy within one standard deviation (Figure~\ref{fig:sched-online}a), although all \texttt{sched\_ext} policies are within one standard deviation of each other there, so these runs show no harm rather than a choice among very different policies. The agentic offline policy loses its lead here, at 2.16$\times$ against 2.19$\times$ for \sys.

Figure~\ref{fig:sched-online}b shows the changing run. \sys changes its setting at most once per stage, within 2.5 seconds of each change. Over the OLTP + batch stage, its p99 latency is 3.9$\times$ below EEVDF, while Long-slice, the setting the policy starts in, lets p99 latency grow to seconds. On the unseen stages, \sys switches to Per-socket while Silo loads its database, before Silo serves requests, and keeps it; its p99 latency stays 3.2$\times$ and 2.1$\times$ below EEVDF, about the same as with Long-slice.

\fbox{%
  \begin{minipage}{\dimexpr\linewidth-2\fboxsep-2\fboxrule\relax}{
      \textbf{Takeaway 5:} \sys's scheduler handles workloads it never saw without harm, and follows a changing workload within one 2-second window.
  } \end{minipage}%
}

\section{Related Work}
\label{sec:related-work}

\heading{Configuration tuning}
Tuners search the parameters that existing policies expose, using Bayesian optimization, reinforcement learning, or LLMs~\cite{alipourfard2017cherrypick,van2017automatic,curino2020mlos,somashekar2024oppertune,chen2024autoos}, including TuxBot~\cite{liargkovas2026tuxbot} (\S\ref{sec:knob-tuning}).
They can only choose among behaviors that a policy already implements. \sys writes the policy itself, and its parameters remain open to such tuners.

\heading{LLM-driven policy generation}
AlphaEvolve-style frameworks such as OpenEvolve and ADRS search over programs offline against an evaluator~\cite{novikov2025alphaevolve,openevolve,cheng2025barbarians} (\S\ref{subsec:evolutionary_search}), and Vulcan synthesizes instance-specialized heuristics but keeps each one fixed at runtime~\cite{vulcan}.
Kgent and SchedCP use agents to synthesize eBPF programs and \texttt{sched\_ext} schedulers~\cite{kgent,zheng2025towards}.
Unlike these, \sys calls an agent only once and adapts the resulting \cacheext~\citep{cacheext} or \texttt{sched\_ext} policy online, without an LLM in the loop.

\heading{Adaptive OS kernel policies}
Classic cache policies adapt to the workload~\citep{megiddo2003arc,216890,264840,211237,yang2023s3fifo}, and LinnOS and LAKE bring learned models into kernel decisions~\citep{linnos2020,LAKE2023}.
Closest to \sys, S4-FIFO separates a simple cache data path from parameter selection, but selects parameters with a model trained offline~\citep{xia2026s4fifo}, whereas \sys searches for them online from measured performance.

\section{Conclusion}
\label{sec:conclusion}
\sys customizes OS policies online by pairing a generalized eBPF kernel policy with a user mode tuner that adjusts its parameters while the workload runs, with no LLM in the loop. Policies that LLM-based evolutionary search or agentic iteration generate offline are competitive on the data they were generated from, but lose much of their performance on unseen data. For the page cache, on data that no approach saw during generation, \sys improves throughput over Linux's LRU by 8.4\%, nearly twice the gain of the best \cacheext policy, at a one-time cost of 16M tokens. For the process scheduler, it improves on EEVDF by 2.75$\times$ across six workloads and follows workload changes within 2.5 seconds. In the cases we evaluated, AI alone did not produce the best policies; the most value came from combining a novel policy design with AI that builds and tunes it.

\section{Acknowledgments}
This work was supported by IBM. We thank the CloudLab team for their assistance.
Tal Zussman was supported by NSF award DGE-2036197 and a Google PhD fellowship.
Georgios Liargkovas was support by Sony.

\bibliographystyle{preprint}
\bibliography{main}

\section*{AI Usage Disclosure}
Generative AI assistants were used to assist with writing and plot generation, among them Claude, Gemini, and ChatGPT.
For writing, this assistance focused on improving style and condensing text, including refining wording and making explanations more concise. For plot generation, AI assistance was used to help develop and refine plotting code and adjust the visual presentation of results.

\clearpage
\appendix
\section{Scheduler Policy and Tuner}
\label{app:sched}

\begin{figure}[h]
  \centering
  \resizebox{0.8\columnwidth}{!}{\begingroup%
\definecolor{archfill}{gray}{0.92}%
\definecolor{archmap}{gray}{0.80}%
\definecolor{archline}{gray}{0.15}%
\definecolor{archmuted}{gray}{0.35}%
\def\fs#1{\fontfamily{phv}\fontsize{\dimexpr #1pt*92/100\relax}%
  {\dimexpr #1pt*110/100\relax}\selectfont}%
\def\note#1{{\fs{7}\color{archmuted}#1}}%
\input{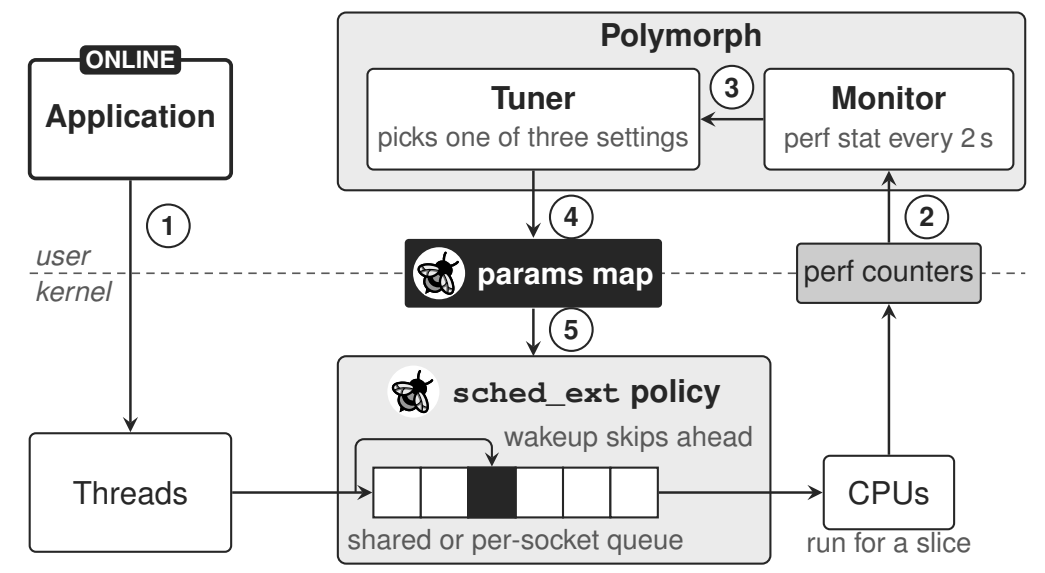}%
\fontfamily{phv}\selectfont%
\resizebox{\linewidth}{!}{%
\begin{tikzpicture}[
  archline, line width=0.5pt, >=stealth,
  container/.style={draw=archline, fill=archfill, rounded corners=2pt},
  comp/.style={draw=archline, fill=white, rounded corners=1.5pt, align=center,
               inner sep=1.5pt, font=\fs{8.2}},
  map/.style={draw=archline, fill=archmap, rounded corners=1pt, align=center,
              inner sep=1pt, minimum height=0.5cm, minimum width=1.55cm, font=\fs{7.6}},
  iface/.style={draw=none, fill=archline, rounded corners=1pt, align=center,
                inner sep=2pt, minimum height=0.5cm, minimum width=1.9cm, text=white,
                font=\fs{7.8}\bfseries},
  task/.style={draw=archline, fill=white, minimum width=0.4cm, minimum height=0.42cm, inner sep=0pt},
  pill/.style={rounded corners=1.5pt, inner sep=1.4pt, fill=archline, text=white,
               font=\fs{6.2}\bfseries},
  sub/.style={font=\fs{7.2}, text=archmuted},
  stepno/.style={circle, draw=archline, fill=white, inner sep=0.4pt,
               minimum size=0.36cm, font=\fs{7}\bfseries},
  flow/.style={->, line width=0.6pt},
]

\draw[archmuted, dash pattern=on 2pt off 1.5pt, line width=0.4pt] (0,2.65) -- (8.4,2.65);
\node[anchor=south west, sub, font=\fs{7.2}\itshape, inner sep=1pt] at (0,2.67) {user};
\node[anchor=north west, sub, font=\fs{7.2}\itshape, inner sep=1pt] at (0,2.63) {kernel};

\node[comp, minimum width=1.7cm, minimum height=1.0cm, line width=0.9pt]
  (in) at (0.85,3.95) {\textbf{Application}};
\node[pill] at (0.85,4.45) {ONLINE};

\draw[container] (2.6,3.35) rectangle (8.4,4.85);
\node[anchor=north, font=\fs{8.2}\bfseries, inner sep=0pt] at (5.5,4.76) {\sys};
\node[comp, minimum width=2.8cm, minimum height=0.85cm] (tun) at (4.25,3.95)
  {\textbf{Tuner}\\\note{picks one of three settings}};
\node[comp, minimum width=2.1cm, minimum height=0.85cm] (mon) at (7.25,3.95)
  {\textbf{Monitor}\\\note{perf stat every 2\,s}};
\draw[flow] (mon) -- (tun);

\node[iface] (params) at (4.25,2.65)
  {\raisebox{\dimexpr -0.5\height+2.5pt\relax}{\ebpfbee}\hspace{3pt}params map};
\node[map] (ctr) at (7.25,2.65) {perf counters};

\node[comp, minimum width=1.7cm, minimum height=1.0cm] (thr) at (0.85,0.8) {Threads};
\draw[container] (2.6,0.2) rectangle (6.25,1.95);
\node[anchor=north, font=\fs{8.2}\bfseries, inner sep=0pt] at (4.425,1.88)
  {\raisebox{\dimexpr -0.5\height+2.5pt\relax}{\ebpfbee}\hspace{3pt}\texttt{sched\_ext} policy};
\foreach \i in {0,...,5}
  \node[task] (q\i) at ({3.1 + 0.4*\i}, 0.8) {};
\node[task, fill=archline] (woken) at (3.9, 0.8) {};
\draw[->, line width=0.5pt, rounded corners=2pt] (2.75,0.8) -- (2.75,1.25) -- (3.9,1.25) -- (woken.north);
\node[sub, anchor=west, inner sep=1pt] at (3.96,1.25) {wakeup skips ahead};
\node[sub] at (4.1,0.38) {shared or per-socket queue};
\node[comp, minimum width=1.1cm, minimum height=0.62cm] (cpu) at (7.25,0.8) {CPUs};
\node[sub] at (7.25,0.38) {run for a slice};
\draw[flow] (q5.east) -- (cpu.west);

\draw[flow] (in) -- (thr);
\draw[flow] (thr) -- (q0);
\draw[flow] (cpu) -- (ctr);
\draw[flow] (ctr) -- (mon);
\draw[flow] (tun.south -| params) -- (params);
\draw[flow] (params) -- (4.25,1.95);

\node[stepno] at (1.17,3.05) {1};
\node[stepno] at (7.57,3.125) {2};
\node[stepno] at (5.925,4.22) {3};
\node[stepno] at (4.57,3.125) {4};
\node[stepno] at (4.57,2.175) {5};

\end{tikzpicture}}%
\endgroup%
}
  \caption{\sys's scheduler. The application's threads~\textcircled{\scriptsize 1} wait in the \texttt{sched\_ext} policy's run queue. A thread that wakes up skips ahead of threads that have waited less than the wake boost (0 or 16\,ms, depending on the setting). Every 2\,s, the monitor reads three perf counters on the workload's CPUs~\textcircled{\scriptsize 2}, and the tuner picks a setting~\textcircled{\scriptsize 3} and writes it into the params map~\textcircled{\scriptsize 4}, which the policy reads at its next decision~\textcircled{\scriptsize 5}.}
  \label{fig:sched-policy-arch}
\end{figure}

Figure~\ref{fig:sched-policy-arch} shows the scheduler policy and tuner evaluated in \S\ref{sec:eval-sched}.

\textbf{Policy.} The policy is a \texttt{sched\_ext} policy with four parameters, kept in a one-entry eBPF array map (Table~\ref{tab:sched-params}). Each scheduling callback reads the map, so a new setting takes effect at the next scheduling decision, without reloading the policy. When a task wakes up and the kernel's default CPU selection finds an idle CPU, the policy places the task directly on that CPU. Otherwise, the task enters a run queue: either one queue shared by all CPUs, or the queue of the socket its CPU belongs to. Each queue is ordered by enqueue time, and a woken task's time is moved earlier by the wake boost. A woken task can therefore pass at most that much waiting work, and no task waits forever. A CPU that needs work takes the first task in its queue. With per-socket queues, it takes a task from another socket's queue when its own is empty. After a change of queue scope, CPUs first drain the queues of the old scope, so no task is left in a queue that is no longer used. The policy keeps separate slices for woken and other tasks, but all three settings use the same value for both. Invalid values fall back to the \texttt{sched\_ext} default (one shared queue, no boost, 20\,ms slices).

\begin{table}[h]
  \centering
  \small
  \caption{Parameters of the scheduler policy and their values in the three settings the tuner chooses from.}
  \label{tab:sched-params}
  \begin{tabular}{@{}llll@{}}
    \toprule
    Parameter & Long-slice & Wake-boost & Per-socket \\
    \midrule
    Queue scope & shared & shared & per socket \\
    Wake boost & 0 & 16\,ms & 0 \\
    Time slice & 20\,ms & 5\,ms & 5\,ms \\
    Idle placement & on & on & on \\
    \bottomrule
  \end{tabular}
\end{table}

\textbf{Tuner.} Every 2 seconds, the tuner reads three counters with \texttt{perf stat} on the workload's 11 CPUs: the \texttt{sched:sched\_wakeup} tracepoint, which counts task wakeups; \texttt{ref-cycles}, which counts reference cycles and stops while a CPU is idle; and \texttt{msr/tsc/}, which counts time-stamp counter cycles and does not stop. From these it computes two statistics:
\begin{denseitemize}
    \item busy CPUs $b = 11 \times \texttt{ref-cycles} / \texttt{tsc}$;
    \item CPU time per wakeup $t = b / w$, where $w$ is the number of wakeups per second.
\end{denseitemize}
It then applies three rules:
\begin{denseenum}
    \item If $b < 2$, the window is a gap between workloads, and the tuner keeps the current setting.
    \item The CPUs are saturated when $b/11 \geq 0.99$. Once saturated, they remain so until $b/11 < 0.95$.
    \item If the CPUs are saturated and $t \geq 400\,\mu$s, the tuner selects Long-slice. If they are saturated and $t < 400\,\mu$s, it selects Wake-boost. Otherwise, it selects Per-socket.
\end{denseenum}
Long bursts on saturated CPUs indicate CPU-bound work, which runs best with long slices. Short bursts on saturated CPUs indicate request threads competing with other work, so woken requests should pass the queue. When there is spare capacity, per-socket queues keep tasks on their socket. The thresholds come from runs of the four seen workloads. The tuner writes to the map only when the setting changes, and it never reads the application's latency or throughput.

\end{document}